\documentclass[12pt]{article}
\usepackage{latexsym}

\usepackage{amsmath, amssymb, amsfonts, amsthm}
\usepackage{bm}

\usepackage[mathscr]{eucal}
\usepackage{url}

\usepackage{xcolor}
\usepackage{tikz}
\usetikzlibrary{shapes.geometric}

\DeclareMathOperator{\argmax}{arg\, max}

\newtheorem{definition}{Definition}

\newcommand \R { \mathbb{R} } 
\newcommand \E { \mathbb{E} } 
\newcommand \N { \mathbb{N} }

\newcommand \mbf {\mathbf}
\newcommand \supp {\mathrm{supp}\,}

\begin{document}

\begin{titlepage}
\begin{center}
\textbf{\large Bounded Rationality with Subjective Evaluations \\
in Enlivened but Truncated Decision Trees%
\footnote{This is a significantly revised version of CRETA working paper \# 89,
which bore the somewhat less informative title 
``Bayesian Rationality 
with Subjective Evaluations in Enlivened Decision Trees''.
A condensed version is due to be published in \textit{Theory and Decision}.}}
\medskip

{{\large Peter J.\ Hammond}: \texttt{p.j.hammond@warwick.ac.uk}} \\ [1ex]
CRETA (Centre for Research in Economic Theory and Applications) \\
Department of Economics, University of Warwick, Coventry CV4 7AL, UK.

\smallskip
2026 January 9th, typeset from \url{ratEnlivenArxiv.tex}       
\end{center}

\medskip
\noindent \textbf{Abstract}:
In normative models a decision-maker is usually assumed 
to be Bayesian rational, and so to maximize subjective expected utility, 
within a complete and correctly specified decision model.
Following the discussion in Hammond (2007) 
of Schumpeter's (1911, 1934) concept of entrepreneurship,
as well as Shackle's (1953) concept of potential surprise, 
we consider enlivened decision trees 
whose growth over time cannot be accurately modelled in full detail.
An enlivened decision tree involves more severe limitations 
than a mis-specified model, unforeseen contingencies, or unawareness,
all of which are typically modelled with reference to a universal state space 
large enough to encompass any decision model that an agent may consider.
We consider a motivating example based on Homer's classic tale 
of Odysseus and the Sirens.
Though our novel framework transcends standard notions of risk or uncertainty,
for finite decision trees that may be truncated because of bounded rationality, 
an extended and refined form of Bayesian rationality is still possible,
with real-valued subjective evaluations instead of consequences 
attached to terminal nodes where truncations occur.
Moreover, these subjective evaluations underlie, for example, 
the kind of Monte Carlo tree search algorithm 
used by recent chess-playing software packages.
They may also help rationalize the contentious precautionary principle.
\texttt{[199 words]}



\end{titlepage}

\section*{Prologue}

\begin{quote}
Grau, teurer Freund, ist alle Theorie; \\ 
Und gr\"un des Lebens gold'ner Baum.%
\footnote{%
One possible translation is: 
``Grey, dear friend, is all theory; and green the golden tree of life.''} \\
--- Mephistopheles in Goethe's \textit{Faust}, Part I.%
\footnote{%
The subject of this paper provided the content 
for my last seminar at Stanford before retiring	in early 2007.
A day or two beforehand, Kenneth Arrow left me a phone message 
asking if I had been inspired by this quotation from Goethe.
While my answer had to be negative, 
I was left feeling that this should have been the source of my inspiration.}
\end{quote}

\begin{quote}
\ldots he said that to finish [the] poem he could not get along without the house 
because down in the cellar there was an Aleph. 
He explained that an Aleph is one of the points in space that contains all other points.

\medskip
The Aleph's diameter was probably little more than an inch,
but all space was there, actual and undiminished.
Each thing (a mirror’s face, let us say) was infinite things, 
since I distinctly saw it from every angle of the universe.
I saw the Aleph from every point and angle,
and in the Aleph I saw the earth
and in the earth the Aleph and in the Aleph the earth;
I saw my own face and my own bowels;
I saw your face; and I felt dizzy and wept,
for my eyes had seen that secret and conjectured object whose name is common to all men
but which no man has looked upon --- the unimaginable universe.
I felt infinite wonder, infinite pity. \ldots \\
for Cantor's Mengenlehre,%
\footnote{%
``Mengenlehre'' is ``set theory'' in German.
}
[Aleph, or $ \aleph $]
is the symbol of transfinite numbers,
of which any part is as great as the whole.

\medskip
Out on the street, going down the stairways inside Constitution Station, riding the subway, every one of the faces seemed familiar to me. I was afraid that not a single thing on earth would ever again surprise me; I was afraid I would never again be free of all I had seen. Happily, after a few sleepless nights, I was visited once more by oblivion. \\
\noindent
--- Excerpts from Jorge Luis Borges \textit{El Aleph} (1945), 
translated by Norman Thomas di Giovanni in collaboration with the author.
\end{quote}

\begin{quote}	
	At the heart of Shackle’s theory of choice is the idea that, 
	when people consider the possible consequences of taking a decision, 
	they give their attention only to those outcomes that they (a) imagine and (b) deem, 
	to some degree, to be possible. 
	This set of outcomes is not guaranteed to include what actually happens, 
	which may be an event they had not even imagined 
	and which comes as a complete surprise to them. \\
\noindent
--- Earl and Littleboy (2014, page 162).
\end{quote}

\section{Background and Outline}
\subsection{Justifying Bayesian Rationality} \label{ss:justBayes}

In decision theory, Bayesian rationality is the hypothesis 
that a decision-making agent makes choices whose consequences, 
which are generally lotteries with both risky and uncertain outcomes, 
maximize the expected value 
of a Bernoulli utility function $Y \owns y \mapsto u(y) \in \R$
defined on a specified non-empty consequence domain $Y$.
For risky consequences which emerge 
from what Anscombe and Aumann (1963) describe as a ``roulette lottery'', 
there is by definition an ``objective'' 
or hypothetical probability $ \pi ( \omega ) \in [0, 1] $
of each lottery outcome $ \omega $ 
in a non-empty finite sample space $ \Omega $.
For uncertain consequences which emerge 
from what Anscombe and Aumann (1963) describe as a ``horse lottery'', 
Bayesian rationality requires there to be 
a ``subjective'' or personal probability $ p(s) \in [0, 1] $ 
of each lottery outcome or state $s$ in a non-empty finite state space $S$.
A general ``Anscombe--Aumann'' lottery specifies, 
for each possible outcome $s \in S$ of a horse lottery 
with subjective probabilities,
a suitable roulette lottery with objective probabilities $ \lambda _s (y) $ 
over consequences $y$ in the non-empty consequence domain $Y$.
Then Bayesian rationality prescribes 
that the appropriate expected utility maximand should be the double sum 
 \( 
 \sum\nolimits _{s \in S} p(s) \sum\nolimits _{y \in Y} \lambda _s (y) u(y) 
 \)
involving products of subjective and objective probabilities.

Past work has offered normative justifications 
for Bayesian rational behaviour in decision trees 
based upon the ``consequentialist'' hypothesis 
set out in Hammond (1988a, b; 1998a, b; 1999). 
This requires the range of possible Anscombe--Aumann consequence lotteries 
which result from prescribed behaviour in any finite decision tree,
including any continuation decision tree,
to be explicable as the value of a suitable choice function defined 
on the relevant domain 
of non-empty finite feasible sets of consequence lotteries.

The more recent justification of Bayesian rationality in Hammond (2022)
replaces consequentialism by the associated concept of ``prerationality''.
This is applied to a family of conditional 
weak base preference relations $ \succsim _E $ given different events $E$.
In fact, prerational preferences require that there exists a behaviour rule 
satisfying the consequentialist hypothesis 
with the property that the associated choice rule selects 
a weakly preferred consequence lottery from each pair of lotteries.

Using either consequentialism or prerationality 
to justify Bayes\-ian rationality
does require one additional well-known continuity axiom.
This axiom applies to preferences over each ``Marschak (1950) triangle'' which, 
for any given triple $ \{ \lambda, \mu, \nu \} $ of roulette lotteries 
of which no two are indifferent, 
is defined as the set $ \Delta ( \{ \lambda, \mu, \nu \} )$ 
of all probability mixtures $ q_\lambda \lambda + q_\mu \mu + q_\nu \mu $ 
of the three lotteries, 
where the three probability weights $ q_\lambda, q_\mu, q_\nu $ 
are all non-negative with $ q_\lambda + q_\mu + q_\nu = 1$. 
After suitable relabelling, 
one can assume that the strict preference relation $ \succ $
satisfies $ \lambda \succ \mu $, $ \lambda \succ \nu $, and $ \mu \succ \nu $.
Then, given the corresponding weak preference relation $ \succsim $,
the continuity axiom requires that both sets
\begin{align*} 	
 & \{ \alpha \in [0, 1] \mid \alpha \lambda + (1 - \alpha ) \nu \succsim \mu \} \\
	\quad \text{and} \quad
 & \{ \alpha \in [0, 1] \mid \mu \succsim \alpha \lambda + (1 - \alpha ) \nu \} 
\end{align*}
should be closed relative to the unit interval $ [0, 1] \subset \R$. 
Dropping continuity would allow some kind of lexicographic preference relation 
over lotteries which is not Bayesian rational.

Some of these earlier papers justifiying Bayesian rationality 
also invoked the assumption of dynamic consistency.
This assumption requires intended or planned behaviour 
at the later decision nodes of a tree $T$ to match actual behaviour.
Yet actual behaviour does not get determined 
at any decision node $n$ of tree~$T$ 
until the decision maker makes a choice 
at the initial node $n$ of the \emph{continuation subtree} $ T_{ \ge n}$ 
which results from eliminating all the nodes of tree $T$ 
that do not weakly succeed~$n$. 
There, by treating~$n$ as the initial node of $ T_{ \ge n}$,
actual behaviour at~$n$ is specified without reference 
to any previous intentions or plans regarding how to behave at~$n$.
So, by considering only actual behaviour at each decision node $n$ of $T$, 
one entirely rules out any possible dynamic inconsistency 
between, on the one hand, actual behaviour at node~$n$ and, on the other hand, 
any previous plans or intentions regarding what to choose at node~$n$. 
In this way, dynamic consistency is satisfied by construction.

\subsection{Bounded Rationality? Or Bounded Modelling?} \label{ss:boundedModels}

 \begin{quote}
 ``All models are wrong, but some are useful.'' \\
 --- George Box (1919--2013)
 \end{quote}

Human ingenuity has led at least some of us to create puzzles 
and other decision problems in order to amuse or instruct each other.
Many children, and some adults, derive satisfaction 
from solving jigsaw puzzles, 
or from learning how not to lose at noughts and crosses, 
otherwise known as tic-tac-toe.
Other people try crossword puzzles, or sudoku, or Rubik's cube.
Generations of students take courses in mathematics 
during which they are expected to learn by solving, 
or understanding the solutions to, progressively more demanding exercises.
In each of these examples the challenge is to find a perfect solution 
to a well defined decision problem.

Typical decision problems, however, 
are not like puzzles or mathematical exercises.
Indeed, they are very often far too challenging 
for full Bayesian rationality to be possible.
This recognition, of course, was a key motivation for Simon (1955, 1957) 
to introduce his concepts of bounded rationality and satisficing.
Yet satisficing seems hard to motivate except as the result of some compromise 
which emerges when the benefits of a more intensive search 
for a Bayesian optimal decision have been traded off 
against the additional cost of that search.
Thus, satisficing seems to apply better to the choice 
of what decision model to analyse
rather than to the choice of what decision to make 
within a given model that is being analysed.
For this reason, it seems that a more satisfactory fundamental concept 
may be that of a bounded model.%
\footnote{I should not claim any originality for this thought.
See, for example, the discussion of Simon, Shackle, and of the game of Chess 
in Section 8.2 of Earl and Littleboy (2014).}
And in the case of decision trees, an obvious form of bounded model
is a truncated tree that results from pruning off 
one or more entire continuation subtrees.

Accordingly, in order for Bayesian rationality to retain any relevance 
for an agent whose behaviour is at best boundedly rational,
the expected utility maximizing model of Bayesian rationality 
needs to be adapted so that it applies even to a truncated decision tree.
This will be the topic of Section \ref{s:truncTrees}.

\subsection{Decision Procedures in Bounded Models} \label{ss:decProced}

Truncation, however, is not the only reason for Bayesian rationality 
in decision trees to be seriously limited in its practical relevance.
Indeed, truncation helps to create an additional reason.
After all, when restricted to one fixed decision tree, 
Bayesian rationality prescribes a single indifference class 
of equally good optimal decision strategies for the agent.
Like a player's strategy in a normal form game, 
at each decision node of the tree,
an optimal strategy will specify an indifference class 
of equally good moves that can be made there.
Furthermore, Bayesian rationality prescribes that the agent should follow,
if not the continuation of an earlier strategy choice,
then an alternative whose expected payoff is the same.

Indeed, an agent who is forced to use a bounded decision model 
ultimately faces the choice, not of a single decision strategy, 
but of what we may call a \emph{decision procedure}.
Such a procedure starts with the agent facing an initial decision node,
where an initial bounded model 
in the form of a truncated decision tree is chosen for analysis.
The procedure continues with the agent finding, and then actually taking,
a decision at the initial decision node which, given the initial bounded model,
is boundedly Bayesian rational.

This initial decision creates the possibility 
that unforeseen later events may occur
which expose some limitations of the initial model.
As the example of Odysseus and the Sirens 
set out in Section \ref{s:homer} illustrates,
if an agent's bounded model excludes significant possibilities,
Bayesian rationality can allow an agent to make bad decisions 
with disastrous consequences.

Alternatively, at any subsequent decision node $n$ 
of the original decision tree $T$,
if all the events that have actually occurred have been foreseen, 
they narrow the range of possibilities 
to a continuation decision tree $ T_{\ge n} $,
which is the part of the tree $T$ that remains relevant 
after reaching node~$n$.
This narrowing will typically release resources which allow an agent 
who contemplates what decision to make at $n$ 
the opportunity to analyse a revised bounded decision tree 
which is richer than $ T_{\ge n} $, the relevant continuation subtree
that had been analysed when making the decision at $n$. 
This, of course, involves an implicit recognition that, 
because of limited resources, the agent had failed to model in full
the consequences of that earlier decision.

Either way, the agent can, and very likely should, 
use an enriched decision model before making the next decision.
The process of reaching that enriched model 
of the relevant continuation subtree, 
and then of making a Bayesian decision at its earliest decision node, 
is the second step of the agent's decision procedure.
Then, of course, unless the end of the decision tree has been reached, 
the decision procedure will continue to a third decision, 
then a fourth, and so on.
The result will be a multi-step decision procedure which cycles between:
\begin{enumerate}
	\item after each decision that is made, reducing the previous decision tree 
	to its continuation subtree given that decision;
	\item enriching that continuation subtree 
	to arrive at a new decision model; 
	\item making a Bayesian rational decision 
	at the initial decision node of that enriched decision tree.
\end{enumerate}
Of course, for sufficiently simplified models, 
the first and third steps of these three are entirely routine;
it is the second step which forms the subject of this paper.

\subsection{Refining Bounded Decision Models} \label{ss:refine}

Myopically following the kind of decision procedure 
described in Section \ref{ss:decProced} may well be unnecessarily inferior.  
This is because, even the bounded decision model 
that the agent uses at any decision node may include an alternative decision
which is superior because it makes some effort 
to anticipate incompletely predictable changes.
After all, if truncation cannot be avoided 
because of computational or other practical modelling limitations,
it will perforce exclude possibilities which should in principle have formed 
part of the agent's original decision tree model.

\medskip

An economic example of this is suggested by the work of Lal (1972).
There is an agent, or agency, who is contemplating a local public project 
such as deepening a village well.
To see if this is worthwhile, the agency might follow Lal 
in applying the well known Little--Mirrlees approach to cost--benefit analysis.%
\footnote{See Little and Mirrlees (1969, 1974, 1991).}
This involves considering not only the obvious costs 
of drilling to make the well deeper, 
and of possibly installing a more powerful pump 
in order to draw water from a greater depth. 
But it should also definitely include the likely general equilibrium effects 
of an increased water supply on local prices in the village
for relevant commodities such as water, agricultural produce, and labour.

After a preliminary cost--benefit analysis has been completed, however,
it may become evident that drawing additional water from the deeper well 
could lower the water table under not only this village, 
but possibly neighbouring villages also. 
This depends on the extent of the hidden aquifer which supplies the deepened well.
Then another decision to consider may be whether to undertake 
a geological groundwater survey intended to estimate how much water 
will be diverted from under neighbouring villages,
and what the general equilibrium effects 
throughout all the affected villages may be.%
\footnote{Yet another decision is whether even to undertake 
an initial cost--benefit analysis, 
as discussed in the appendix to Little and Mirrlees (1991).}
In the end, what may have seemed initially like a simple decision problem 
could well become greatly complicated 
by considerations that were initially unforeseen.

\medskip

A second example concerns the game of Chess. 
Suppose there were an unboundedly rational chess player, 
or chess-playing algorithm. 
By definition, this player could always make all the calculations 
needed to evaluate any decision tree perfectly, 
and so would always play a perfect game.
In practice, however, any human player, 
and even the strongest chess playing computer algorithms,
will be able to analyse in full
only a small proportion of all the possible game continuations.
Thus, any practicable analysis will typically have to be truncated 
well before the result of the game has become indisputable.
In this sense, a player will nearly always have to use 
a bounded decision model. 

Even so, after a preliminary analysis, 
usually a skilled player will be able to identify 
not only which possible continuations are most likely to occur, 
but also which are critical in the sense of being most likely 
to change the outcome of the game.
Then the player may decide, before choosing the next move, 
which of these likely and/or critical continuations 
deserve further analysis in an enriched model of the game.

Considerations of this kind suggest that there can be many instances
where a rational agent should want to refine 
whatever bounded model has been used
to analyse the first decision in a tree even before making that decision.
Even more obvious is the probable benefit at any subsequent decision node 
of refining whatever decision model had been used 
to analyse decisions at previous decision nodes.

\subsection{Enlivened but Truncated Decision Trees} \label{ss:introEnliv}

Accordingly, in order to allow the decision tree to change, even unpredictably,
a framework with ``enlivened'' decision trees is proposed.
An entirely myopic agent who follows the old adage 
``Don't cross your bridges before you come to them'' 
--- which Savage (1963, p.\ 16) in particular has discussed
--- will act as though this enlivening is totally irrelevant. 
This leads to the agent lurching from one model to the next, 
displaying hubris throughout.

Of course many future decisions and their uncertain consequences 
cannot be modelled in any detail. 
Nevertheless, an agent with even a little sophistication should recognize 
that what matters for any one decision is the current expectation of what,
when viewed in retrospect, its ultimate \textit{ex post} value will be. 
Following ideas that Koopmans (1964) and Kreps (1990, 1992) developed 
in order to discuss the preference for flexibility, an agent should seek 
to determine these expected valuations as reliably as possible, 
using whatever limited evidence is deemed to be relevant, 
as well as what can be handled within whatever bounded resources 
the agent can afford to allocate to decision analysis.
See also Dekel \textit{et al.} (2001, 2005) and many successors 
for related ideas in the context of decision making 
with unforeseen contingencies whose possibility is, nevertheless, 
foreseen by an apparently omniscient and hubristic decision analyst.
The example of the game of Chess, which is discussed more fully 
in the fuller working paper version of this article, 
illustrates how unreasonable it can be to postulate 
the existence of such an omniscient decision analyst.%
\footnote{See also the related work on unawareness 
in decisions and games by, inter alia, Schipper (2014a, b), 
Halpern and R\^ego (2014), Grant \textit{et al.} (2015a, b),
and especially the related work on growing awareness and reverse Bayesianism 
by Vier\o\ (2009, 2021) and by Karni and Vier\o\ (2013, 2015, 2017).
Other relevant work on unawareness includes the papers published 
in the special issue of \textit{Mathematical Social Sciences} 
edited by Schipper (2014a), as well as those cited in Vier\o \ (2021).}

We emphasize that the present paper differs from this earlier work 
on unforeseen contingencies or unawareness 
by not relying on the existence of any ``augmented conceivable state space''
of the kind defined in Karni and Vier\o\ (2017, p.\ 304).
Instead, we allow the relevant state space to grow entirely unpredictably 
as a result of the dynamic process that we call ``enlivenment''.
Specifically, though a decision-making agent may be aware 
of the possibility of their own unawareness, 
they are unable even to formulate a practical model which is based, 
as usual, on a comprehensive space of all conceivably possible states.
This enrichment of the previous concepts 
of unforeseen contingencies or unawareness,
which was introduced informally in Hammond (2007), 
is inspired in part by Schumpeter's (1911, 1934) concept of entrepreneurship,
as well as by Shackle's (1953) concept of potential surprise.%
\footnote{See Metcalfe \textit{et al.}\ (2021) for a recent discussion 
of links between these two concepts.}
The concept of an enlivened decision tree can be motivated in part 
by the classical example of Odysseus and the Sirens 
discussed in Section~\ref{s:homer}.

That said, postulating a fully enlivened decision tree 
that is never subject to further enlivenment 
raises the same kind of conceptual problem as that associated 
with assuming the existence of a universal augmented conceivable state space.
To avoid this issue, we recognize the relevance 
of recursively enlivened decision trees 
whose growth can never be fully described in a single universal model.
Nevertheless, we argue that even a fully enlivened tree 
can and should still be reduced to one with random outcomes which, 
instead of consequence lotteries, are subjective evaluations 
attached to the terminal nodes of a truncated decision tree.

\subsection{Outline of Paper}

Earlier work that was briefly reviewed in Section \ref{ss:justBayes} 
shows how consequentialism or prerationality, 
when supplemented by a continuity condition, 
offers a normative justification for Bayesian rational behaviour 
within any given decision tree $T$.
Our task in this paper is to show how to modify 
the postulate of Bayesian rationality so it can become a decision procedure 
that allows the agent to adapt to a decision tree 
that is both evolving because of enlivenment,
as well as subject to the inevitable truncation which is needed 
to avoid complications that are too deep to be analysed in full.

Section \ref{s:ubvbr} begins the rest of the paper by offering a brief review 
of some distinctions between unbounded and bounded rationality.
These include prominent examples 
such as Simon's concept of procedural rationality,
as well as Manzini and Mariotti's (2007) ``rational shortlist'' method.

Next, Section \ref{s:homer} revisits the well known Homeric example 
of Odysseus and the Sirens.
Previous work 
such as Strotz (1956), Pollak (1968), Hammond (1976) and Elster (1979)
has typically regarded this as a prominent example of changing tastes,
illustrating the distinction between na\"\i ve and sophisticated choice,
as well as the potential value of commitment devices.
Here, by contrast, this Homeric example is viewed as a mythical decision tree
which the sorceress Kirke (or Circe) enlivened as the sage advice 
that she was offering Odysseus progressed through several stages.%
\footnote{%
In Homeric Greek, 
the spelling of her name is $K \acute { \iota } \rho \kappa \eta $.
}

The next two Sections \ref{s:linQuad} and \ref{s:chess} 
focus on two more particular examples. 
The first is of a consumer who, as an investor, chooses a portfolio of financial assets
in order to maximize a two-period quadratic utility function 
subject to a linear budget constraint.
Enlivening this consumer's decision problem could merely affect parameter values,
but it could also allow the possibility that new commodities, 
which may even not yet have been invented, could become relevant.

In Section \ref{s:chess}, the second of these two particular examples concerns the game of Chess,
whether played by computers or by humans. 
Of course, as a two-person game, superficially Chess goes beyond the single-person decision problems
that are the main subject of the paper.
Nevertheless, somebody who analyses a Chess game, whether played in the past or taking place currently,
faces the task of trying to find optimal play by both players.%
\footnote{Footnote 12 in Section \ref{ss:ubvbr} describes Chess positions which can be analysed perfectly.}
In any case, Chess exemplifies rather dramatically 
the importance of a decision maker's bounded model, whether that model is used 
in a single-person decision problem or a two-person game. 

Section \ref{s:chess} offers a cursory explanation 
of how Monte Carlo tree simulation can allow a computer algorithm to evaluate positions 
that arise after possible future moves have been analysed in detail as far as possible. 
It also presents a brief case study of a particularly unfortunate human move, 
described at the time as ``the blunder of the century''. 
This is seen as one particularly prominent player's failure 
to revise his bounded model of how the game was likely to proceed.

Section~\ref{s:bayesRat} provides a summary of the key concepts we need
to describe Bayesian rational behaviour in classical finite decision trees
that are neither truncated nor enlivened.
A key tool used in later analysis 
is the evaluation $ v(T) $ of any decision tree~$T$.
This is defined as the normalized expected utility generated 
by any consequence lottery that can result from deciding optimally 
at each decision node of $T$.
Following the principle of optimality in dynamic programming, 
this evaluation can be calculated by backward recursion, 
starting at the terminal nodes of $T$, 
each of which has a specified consequence 
that generally takes the form of a lottery.

As was discussed briefly in Section \ref{ss:refine}, any agent 
contemplating a decision whose consequences are too hard to analyse in full
will have to resort to a bounded model.
Section~\ref{s:truncTrees} introduces 
a particular but especially important kind of bounded decision model
in which many continuation subtrees get cut off.
The result is a ``truncated'' decision tree with consequences 
which are specified initially only for terminal nodes of a tree 
that truncation has not removed.
Then the backward recursion method of evaluation 
used in Section~\ref{s:bayesRat} fails for those nodes 
of a truncated decision tree that may eventually be succeeded 
by ``truncation'' nodes 
rather than by terminal nodes of the original complete tree.
Section~\ref{ss:menuConseq} proposes filling this gap 
by assigning to each truncation node a ``subjective evaluation'' 
in the form of a real-valued estimate 
of what the evaluation of the missing continuation subtree starting there 
would have been if a complete analysis of that subtree were possible.

Section \ref{s:enlivNodes} turns at last to describing more formally
how decision trees become enlivened.
We consider a process that has been broken down into a sequence of steps.
Each of these steps is a special kind of ``minimal'' enlivenment 
of the previous tree in the process that occurs 
either at one specified terminal node or else along one specified edge.
The whole process starts with a given ``base'' decision tree $T$
and results finally in an enlivened decision tree $ T^+ $.

Next, Section \ref{s:ratEnliv} explains how to extend 
the concept of a Bayesian rational decision procedure
to the kind of enlivened but truncated decision tree 
which a boundedly rational agent may use.
A key part of this extension recognizes that, 
as in example of Odysseus and the Sirens discussed in Section~\ref{s:homer},
enlivenment may involve extending 
the domains of consequences and of states of the world, 
and so those of the enlivened utility functions 
and subjective probability distributions. 
Section \ref{s:ratEnliv} concludes by comparing 
the arbitrariness of utilities and subjective probabilities
in our model of Bayesian rationality with enlivened subjective evaluations
to the arbitrariness of the corresponding concepts 
in the Anscombe and Aumann (1963) model of subjective probability.

The penultimate Section \ref{s:apps} starts in Section \ref{ss:precPrinc}
with some discussion of the precautionary principle 
and its relation to option values.
Next, Section \ref{ss:transExpers} offers a brief discussion of recent work 
by Ullmann-Margalit (2006), Paul (2014, 2015a, b, c) and other philosophers 
who have introduced the concept of a ``transformative experience''.
Section \ref{ss:menuConseqs} briefly discusses 
how to extend the results of this paper
to the framework in Hammond and Troccoli Moretti (2025)
where decision trees may have non-terminal timed consequence nodes.
These include nodes with ``menu consequences'' which depend 
on the continuation subtree whose initial node is the consequence node.
Last, Section \ref{ss:revBayes} discusses recent work 
concerning the concept of ``reverse Bayesianism'' 
due to Karni and Vier\o\ (2013, 2015, 2017).

The brief concluding summary that makes up Section \ref{s:concs} 
focuses on describing a bounded Bayesian rational decision procedure
which can be applied to enlivened but truncated decision trees.

\section{Beyond Unbounded Rationality} \label{s:ubvbr}

\begin{quote}
To see a World in a Grain of Sand \\
And a Heaven in a Wild Flower, \\
Hold Infinity in the palm of your hand \\
And Eternity in an hour. \\
--- From William Blake's ``Auguries of Innocence''
\end{quote}

\subsection{Unbounded versus Procedural Rationality} \label{ss:ubvbr}

Simon's (1955, 1957) famous concept of ``bounded rationality''
may perhaps best be defined by its negation.
Decision agents who are \emph{unboundedly rational} make perfect decisions
based on perfect models of all the possible acts they could choose,
along with all their potential consequences.
The result could be the rather disturbing kind of hypothetical complete model 
so artfully described in Jorges Luis Borges' short story ``El Aleph'',
from which some relevant extracts are quoted in the prologue to this paper.

The definition of unbounded rationality in perfect models remains the same 
no matter whether the consequences are certain (determinate),
or else, using the terminology due to Anscombe and Aumann (1963): 
(i) risky, 
with hypothetical ``objective'' probabilities as in a roulette lottery;
(ii) uncertain, 
with personal or ``subjective'' probabilities as in a horse lottery.
Such unbounded rationality would threaten 
to make games as complicated and enthralling as Chess or Go
no more interesting than the children's game of noughts and crosses, 
also known as ``tic-tac-toe''.%
\footnote{%
Note that in Chess, the ``Lomonosov tablebases'' 
that are distributed online at \url{http://tb7.chessok.com/}
currently describe perfect play starting from any legally possible position 
\emph{provided} that there is a total of no more than seven pieces 
of either colour left on the board, including both kings.
The usual game of Chess starts, of course, 
with each of the two players having 16 pieces on the board.}
And there would be no such thing as the ``law of unintended consequences'';
every possible consequence should be calculated, 
making it in some sense intentional, 
even as the perhaps unfortunate outcome of a risky decision.

In addition to bounded rationality,
Simon advanced the important related idea of ``procedural rationality''.
This recognizes that decision \emph{procedures} could be rational,
even if they lead to decisions that are irrational 
in the sense of violating unbounded rationality.
He emphasized concepts like \emph{aspiration level}, 
along with \emph{satisficing}.
The latter appears to mean finding a decision 
that reaches the aspiration level, 
and making a decision that seems merely good enough, rather than optimal.
But optimal (or even just flexible) search suggests
that if the aspiration level is reached quickly and easily,
it is too undemanding and so should be raised.

The normative framework this paper proposes, by contrast, 
suggests that satisficing behaviour should occur, 
not within a given decision model, but in choosing how much detail 
to include in the model and how much to exclude from it.
Then ultimately behaviour should be optimal 
relative to whatever bounded model has been selected for analysis.

\subsection{Rational Shortlists and Other Bounded Models} \label{ss:ratShortlist}

One kind of bounded model involves the ``rational shortlist'' 
that Manzini and Mariotti (2007) introduced 
to discuss their concept of ``sequential rationalizability''
for the case of decision problems whose acts 
have only determinate consequences.
Their idea is that, given an incoveniently large feasible set of options,
at an initial stage the agent could shortlist
a relatively small subset for later serious consideration.
Moreover, this shortlist should be small enough to make finding 
a fully optimal decision amongst those that are shortlisted 
a manageable decision problem.
Thus, any shortlist can be thought of as a bounded model of the feasible set.
Also, when it is recognized that observation and/or computation can be costly, 
work on ``rational inattention'' inspired by Sims (2003, 2011) 
and by Hansen and Sargent (2007) 
considers what bounded decision model may be optimal.

The choice of shortlist can be supposed to emerge rather arbitrarily, 
even randomly, from some kind of boundedly rational search procedure.
Of course, some options may be much more likely to be shortlisted than others.
Also, if the composition of the shortlist is regarded as random,
the different random variables indicating whether each option 
belongs to the shortlist may well be correlated.

Once the shortlist has been determined at the first stage, however,
it is entirely reasonable to assume that, at a subsequent second stage,  
the agent indeed selects an optimal element 
among those that have been shortlisted.
That is, choice from within the shortlist satisfies 
what Simon (1955, 1957) would call ``substantitive rationality''.

Shortlisting can be viewed as a particular form of procedural rationality,
involving a two-stage procedure.
The main point to be made here, however, is that whatever the shortlist may be,
it represents a \emph{bounded} model of the full decision problem.
Indeed, limitations like the inability of computers to play chess perfectly
apply to all difficult decision problems, 
including most of those that arise in life 
rather than in the oversimplified models 
that are typically analysed and applied
by economists and other decision scientists. 
For this reason, any model we use to inform our decision-making
should be flexible enough to allow graceful adaptation to potential changes
that any practical model must otherwise ignore.

Suppose an effort really is made 
to take Simon's ``procedural rationality'' idea as seriously as possible.
Specifically, it is presumably interesting to explore 
the implications of assuming that, for decision-making agents:
\begin{enumerate}
     \item their time, attention, and computational resources
     are far too limited for all but simplified models;
     \item and in fact they confine themselves to bounded models
     which are sufficiently simple that they really can find
     the decision that is optimal within the confines of their bounded model.
\end{enumerate}

Once one recognizes, however, that the model 
which an agent uses for making decisions is bounded, 
then one must also recognize that events may eventually force consideration 
of an expanded or ``enlivened'' model that includes unmodelled changes.

\section{Odysseus and the Sirens Revisited} \label{s:homer}
\subsection{A Na\"\i ve Sailor's Model}

As our first ``classical'' example of an enlivened decision tree, 
we reconsider the Homeric myth of Odysseus and the Sirens.
According to this epic myth, na\"\i ve sailors whose shortest sea route
passed near the Sirens' island had perhaps in the past 
used a bounded model of their decision tree
like the one illustrated in Figure~\ref{f:naive}.%
\footnote{Note that in Figures \ref{f:naive}--\ref{f:odyss2},
each square box indicates a terminal node 
with a consequence described by the word in the box.
Apart from these terminal nodes, all other nodes of the tree are decision nodes.
Apart from the initial decision node labelled ``start'', 
each other decision node is labelled with a word or phrase
describing very briefly what is the last action that would lead to that node. 
}
Specifically, these na\"\i ve sailors acted as though 
they thought that their choice was between:
\begin{itemize}	
\item either {\color{brown}going near} the Sirens' island 
and reaching their destination {\color{brown}early} by a direct route;
\item or {\color{orange}avoiding} the Sirens' island and arriving {\color{orange}late} 
after a detour.
\end{itemize}

\begin{figure}[htb]
	\hfill
\begin{minipage}[t]{5cm}
\begin{tikzpicture}
\node[ellipse,draw] {start}[grow'=up]
	child[orange] {node {avoid} 
		child {node [rectangle,draw] {late}}
		child[missing] 
		} 
	child[brown] {node {go near} 
		child {node [rectangle,draw] {early}
		}};
\end{tikzpicture}
\caption{Na\"\i ve Sailor} \label{f:naive}
\end{minipage}
\hfill
\begin{minipage}[t]{7cm}
\begin{tikzpicture} 
\node [ellipse,draw] {start}[grow'=up]
	child[orange] {node {avoid} 
		child {node [rectangle,draw] {late}}
		child[missing] 
		}
	child[brown] {node {go near}
		child {node {go on} 
			child {node [rectangle,draw] {early}}
			}
	child[red] {node {tarry} 
			child {node [rectangle,draw] {die}} 
			}};
\end{tikzpicture}
\caption{Sophisticated Sailor} \label{f:soph}
\end{minipage}
\hfill
\end{figure}

\subsection{A Sophisticated Sailor's Model}

According to Homer, however, Odysseus has the sorceress Kirke as a supernaturally well-informed adviser.
She warned Odysseus that the Sirens' singing had the power 
to lure unwary sailors to their deaths,
and that the meadows on the Sirens' island were littered with sailors' bones.
So if any na\"\i ve sailor came within earshot of the Sirens 
by choosing {\color{brown}go near} in the decision tree shown in Figure~\ref{f:naive},
they would find themselves facing instead the decision node marked {\color{brown}go near} 
in the enlivened and so expanded decision tree shown in Figure~\ref{f:soph}.
At this node in the enlivened tree, their apparent choice would be:
\begin{itemize}	
\item either {\color{brown}go on} home after hearing the Sirens, 
\item or {\color{red}tarry}, enchanted by their singing,
	and {\color{red}die} on their island, 
	before ever reaching their intended destination.
\end{itemize}
Of course, the added feature was that, after hearing the Sirens, 
no previous sailor had ever exercised enough will-power to escape the island.
This is the essential characteristic of what, in Hammond (1976), 
was called ``potential addict'' example of changing tastes.
Faced with the decision tree of Figure~\ref{f:soph}, 
a sophisticated sailor who understands the persuasive power of the Sirens' singing 
would avoid their island and stay out of earshot,
even at the cost of only reaching their intended destination after a significant delay.

\subsection{Kirke's First Enlivened Model for Odysseus}

\begin{figure}[htb]
	\hfill
\begin{minipage}[t]{6cm}
\begin{tikzpicture} 
\node [ellipse,draw] {start}[grow'=up]
	child[orange] {node {avoid}
		child {node [rectangle,draw] {late}}
		child[missing] 
		}
	child[brown] {node {go near} 
		child[blue] {node {wax} 
			child {node [rectangle,draw] {early}}
			child[missing] 
		}
		child {node {no wax}
			child {node {go on} 
				child {node [rectangle,draw] {early}}
				}
			child[red] {node {tarry} 
				child {node [rectangle,draw] {die}} 
				}}};
\end{tikzpicture}
\caption{Kirke's First Model} \label{f:odyss1}
\end{minipage}
\hfill
\begin{minipage}[t]{7cm}
\begin{tikzpicture} 
\node [ellipse,draw] {start}[grow'=up]
	child[orange] {node {avoid}
		child {node [rectangle,draw] {late}}
		child[missing] 
		}
	child[brown] {node {go near} 
		child[blue] {node {wax} 
			child[violet] {node {bind}
				child {node [rectangle,draw] {hear}}
				}
			child {node {free}
				child {node [rectangle,draw] {early}}
			}
			child[missing] 
			}
		child {node {no wax}
			child[missing] 
			child {node {go on} 
				child {node [rectangle,draw] {early}}
				}
			child[red] {node {tarry} 
				child {node [rectangle,draw] {die}} 
				}}};
\end{tikzpicture}
\caption{Kirke's Final Model} \label{f:odyss2}
\end{minipage}
\hfill
\end{figure}

Kirke's advice was not confined to a warning, however.
Rather routine and unheroic stories about avoiding the Sirens' island 
and getting back to Ithaca somewhat late by a roundabout route 
do not constitute memorable epics.
Instead Kirke drew attention to the possibility of sailing safely past the Sirens' island,
provided the precaution was taken of stopping up all the sailors' ears with wax.
Thus, after deciding to approach the Sirens' island, 
but before getting within earshot, 
the choice at the node {\color{brown}go near} in Figure~\ref{f:odyss1} would be:
\begin{itemize}	
\item either {\color{blue}wax} all the crew's ears (including those of Odysseus himself),
so none of them hears the Sirens;
\item or use {\color{brown}no wax}, like all the earlier na\"\i ve sailors
whose bones now litter the Sirens' meadow.
\end{itemize}

\subsection{Kirke's Final Enlivened Model for Odysseus}

A much more interesting epic, however, is the one that Homer has given us.
Homer had Kirke advise Odysseus on an even better course of action 
which allowed Odysseus, at least, to hear the Sirens and yet escape with his life.
Indeed, Odysseus was advised that, 
in addition to arranging for the ears of all his crew to be waxed,
he should have himself bound tightly to the mast.
Also, his crew should be given strict instructions that, 
in response to any pleas for release that they see Odysseus making,
not only should these pleas be ignored,
but also the tightness of his bounds should be increased even more.
Thus, Kirke's  final model for Odysseus includes an extra decision node
marked {\color{blue}wax} in Figure~\ref{f:odyss2}, where the choice is between:
\begin{itemize}	
    \item either {\color{violet}binding} Odysseus to the mast,
    with ears unwaxed so he can {\color{violet}hear} the Sirens, 
    \item or leaving Odysseus {\color{blue}free},
    but with ears waxed like the rest of the crew.
\end{itemize}
We emphasize that this extra decision can be made after steering toward the Sirens' island,
but it \emph{must} be made before getting close enough to hear their singing.

\subsection{Toward Enlivened Decision Trees}

The earlier na\"\i ve sailors whose bones littered the Sirens' meadow
had a model like that in Figure~\ref{f:naive}.
Once they had heard the Sirens' singing and so learned of their existence,
they may have realized too late that a more appropriate model 
would have been like that in Figure~\ref{f:soph}.
Odysseus (and his crew) were fortunate enough to be provided
with a much more useful model, 
going even beyond Figure~\ref{f:soph} to Figure~\ref{f:odyss1} in the first instance,
then ultimately to Figure~\ref{f:odyss2}.

Each decision tree in Figures~\ref{f:naive}--\ref{f:odyss2}
is lifeless when considered in isolation.
The four trees together, however, tell an epic tale of learning.
But it is \emph{not} the usual statistical model of learning more and more 
about the state of the world within a fixed sample space.
Rather, the set of possibilities is expanding, 
as more and more possibilities are included in the enriched model.
By introducing the term ``enlivened tree'', 
I have not resisted the temptation to draw an analogy with a live growing tree.
Nor of suggesting a strong analogy to the works of Schumpeter (1911, 1934) on innovation,
and of Shackle (1953) on ``potential surprise''
--- see Hammond (2007) for further discussion.

\section{A Linear--Quadratic Portfolio Problem} \label{s:linQuad}
\subsection{A Two-Period Portfolio Problem} \label{ss:baseProblem} 

The first of our two extended examples concerns a consumer 
with a two-period Bernoulli utility function that takes the quadratic form 
\begin{equation} \label{eq:bernUtil}
	u( \mbf x_1, \mbf x_2 ) =
	- \tfrac 12 ( \mbf x_1 - \mbf a_1 )^\top \mbf Q_1 ( \mbf x_1 - \mbf a_1 ) 
	- \tfrac 12 ( \mbf x_2 - \mbf a_2 )^\top \mbf Q_2 ( \mbf x_2 - \mbf a_2 ) 
\end{equation}
Here $ \mbf x_1 $ and $ \mbf x_2 $ denote finite-dimensional consumption vectors in the two periods, 
which may possibly have different dimensions,
whereas $ \mbf a_1 $ and $ \mbf a_2 $ are corresponding parameter vectors.
Furthermore, assume that $ \mbf Q_1 $, $ \mbf Q_2 $ 
are symmetric and positive definite square matrices of appropriate dimension.

Suppose that the consumer faces two budget constraints, one each period, which can be written as
\begin{equation} \label{eq:budgEqa}
		\mbf p_1^\top \mbf x_1 + \mbf q^\top \mbf b = m_1 \quad \mbox{and} \quad
	\mbf p_2^\top \mbf x_2 = m_2 + \mbf r^\top \mbf b 
\end{equation}
where $ \mbf b$ denotes a finite-dimensional portfolio vector 
of net asset holdings at the end of period 1,
with $ \mbf q$ as the asset price vector in period 1, 
and then $ \mbf r$ as the gross return vector.
Of course $ \mbf p_1 $ and $ \mbf p_2 $ denote commodity price vectors each period,
both assumed to be strictly positive, whereas $ m_1, m_2 \in \R $ are outside wealth transfers.
We allow $ \mbf a_2 $, $ \mbf r$ and $ m_2 $ all to be uncertain, 
but treat $ \mbf p_2 $ as certain,
just as Hicks (1946) did when he used point expectations of future prices
in his theory of temporary equilibrium.%
\footnote{For a somewhat similar idea, see Myerson (1983).}
For simplicity we also assume that the symmetric matrix $ \mbf Q_2 $ is known in period 1.
Finally, we assume that the random gross return vector~$ \mbf r$ 
is stochastically independent of both random variables $ \mbf a_2 $ and $ m_2 $.   

\subsection{The Second-Period Optimum}

By the start of period 2, we assume that the parameter vector $ \mbf a_2 $, 
the gross return vector $ \mbf r$, and unearned income $ m_2 $ have all become known, 
along with the portfolio vector $ \mbf b$ 
which is pre-determined by the consumer's own choice in period 1,
Accordingly, the consumer's second-period optimization problem, 
which is independent of whatever $ \mbf x_1 $ is chosen in period 1, reduces to
\begin{equation} \label{eq:per2}
	\max _{ \mbf x_2 } \,
	\left\{ - \tfrac 12 ( \mbf x_2 - \mbf a_2 )^\top \mbf Q_2 ( \mbf x_2 - \mbf a_2 ) \right\}
	\quad \mbox{subject to} \quad \mbf p_2^\top \mbf x_2 = m_2 + \mbf r^\top \mbf b 
\end{equation}
To solve this constrained maximization problem, introduce the Lagrangian
\begin{equation} \label{eq:Lagrange}
	\mathcal L_{ \lambda _2 }( \mbf x_2 )
 = - \tfrac 12 ( \mbf x_2 - \mbf a_2 )^\top \mbf Q_2 ( \mbf x_2 - \mbf a_2 ) 
 - \lambda _2 ( \mbf p_2^\top \mbf x_2 - m_2 - \mbf r^\top \mbf b) 
\end{equation}
Then $ \mathcal L_{ \lambda _2 }( \mbf x_2 )$ is concave as a function of $ \mbf x_2 $.
So it is maximized at any point $ \mbf x_2 $ that satisfies the first-order condition
\begin{equation} \label{eq:FOC}
	\mbf 0 = \mathcal L'_{ \lambda _2 } ( \mbf x_2 ) 
 = - ( \mbf x_2 - \mbf a_2 )^\top \mbf Q_2 - \lambda _2 \mbf p_2^\top 
\end{equation}
Because $ \mbf Q_2 $ is assumed to be positive definite and so invertible, 
this first-order condition is evidently equivalent to
\begin{equation} \label{eq:FOC2}
	( \mbf x_2 - \mbf a_2 )^\top = - \lambda _2 \mbf p_2^\top \mbf Q_2 ^{-1} 
\end{equation} 
or, after transposing and rearranging, to
\begin{equation} \label{eq:x2}
	\mbf x_2 = \mbf a_2 - \lambda _2 \mbf Q_2 ^{-1} \mbf p_2 \end{equation}
Substituting this into the budget equation in (\ref{eq:per2}) gives
\begin{equation} \label{eq:FOC3}
	\mbf p_2^\top \mbf x_2
	= \mbf p_2^\top ( \mbf a_2 - \lambda _2 \mbf Q_2 ^{-1} \mbf p_2 )
	= m_2 + \mbf r^\top \mbf b \end{equation}
implying that 
\begin{equation} \label{eq:lambda2}
	\lambda _2 = \frac { \mbf p_2^\top \mbf a_2 - m_2 - \mbf r^\top \mbf b}
	{ \mbf p_2^\top \mbf Q_2 ^{-1} \mbf p_2 } 
\end{equation}
Note that the solution $ \lambda _2 $ exists 
because $ \mbf p_2 \ne \mbf 0$ and $ \mbf Q_2 $ is positive definite.
Finally, we can combine \eqref{eq:lambda2} with \eqref{eq:x2} 
to determine the optimal demand vector, which is
\begin{equation} \label{eq:optDemand}
		\mbf x^*_2 = \mbf a_2 - \frac { \mbf p_2^\top \mbf a_2 - m_2 - \mbf r^\top \mbf b}
	{ \mbf p_2^\top \mbf Q_2 ^{-1} \mbf p_2 } \mbf Q_2 ^{-1} \mbf p_2 
\end{equation}
Of course, for this solution to be economically sensible, 
we should require that $ \lambda _2 \ge 0$,
or equivalently, that $ \mbf p_2^\top \mbf a_2 \ge m_2 + \mbf r^\top \mbf b$. 
Because this inequality involves the asset vector $ \mbf b$ chosen in the first period,
we will return to this issue later 
after deriving the consumer's optimal decisions in the first period.

Note that this solution implies that \textit{ex post},
after $ \mbf a_2 $, $ \mbf r$ and $ m_2 $ have all become known 
and $ \mbf x^*_2 $ has been chosen optimally, equations \eqref{eq:x2} and \eqref{eq:lambda2} 
imply that the consumer's maximized second period utility is
\begin{align}
	- \tfrac 12 ( \mbf x^*_2 - \mbf a_2 )^\top \mbf Q_2 ( \mbf x^*_2 - \mbf a_2 ) 
	&= - \tfrac 12 \lambda _2^2 \mbf p_2^\top \mbf Q_2 ^{-1} \mbf Q_2 \mbf Q_2 ^{-1} \mbf p_2 \notag \\
	&= - \frac {( \mbf p_2^\top \mbf a_2 - m_2 - \mbf r^\top \mbf b) ^2}
			{2 \mbf p_2^\top \mbf Q_2 ^{-1} \mbf p_2 } \label{eq:util2} 
\end{align}

\subsection{First-Period Expected Utility}

Coming back to the first period,
we have assumed that $ \mbf p_2 $ and $ \mbf Q_2 $ are both known in advance.
So after using \eqref{eq:util2}, the \emph{ex ante} expected value 
of the intertemporal Bernoulli utility function \eqref{eq:bernUtil} can be expressed as the function
\begin{equation} \label{eq:nmuf1}
	v( \mbf x_1, \mbf b) 
	= - \tfrac 12 ( \mbf x_1 - \mbf a_1 )^\top \mbf Q_1 ( \mbf x_1 - \mbf a_1 ) 
 	- \frac { \E ( \mbf p_2^\top \mbf a_2 - m_2 - \mbf r^\top \mbf b) ^2}
			{2 \mbf p_2^\top \mbf Q_2 ^{-1} \mbf p_2 } 
\end{equation}
of the first-period choice variables $ \mbf x_1 $ and $ \mbf b$.
The numerator of the fraction in the second term of the right-hand side of \eqref{eq:nmuf1}
can be expanded as
\begin{multline} \label{eq:expsq}
	 \E ( \mbf p_2^\top \mbf a_2 - m_2 - \mbf r^\top \mbf b) ^2 \\
	= \E ( \mbf p_2^\top \mbf a_2 - m_2 ) ^2
	- 2 \E [( \mbf p_2^\top \mbf a_2 - m_2 )( \mbf r^\top \mbf b)]
	+ \E ( \mbf r^\top \mbf b) ^2
\end{multline}
Let $ \bar { \mbf a}_2 := \E \mbf a_2 $, $ \bar m_2 := \E m_2 $ 
and $ \bar { \mbf r} := \E \mbf r$ denote the respective means, 
all of which are assumed to exist.
Our assumption that $ \mbf r$ is stochastically independent of $ \mbf a_2 $ and $ m_2 $ 
implies that the middle term on the right-hand side of (\ref{eq:expsq}) reduces to
\begin{equation} \label{eq:corr}
	\E [( \mbf p_2^\top \mbf a_2 - m_2 )( \mbf r^\top \mbf b)] 
	= ( \mbf p_2^\top \bar {\mbf a}_2 - \bar m_2 )( \bar { \mbf r}^\top \mbf b)
\end{equation}

As for the last term on the right-hand side of (\ref{eq:expsq}), note that 
\begin{equation} \label{eq:rbsq}
	 ( \mbf r^\top \mbf b) ^2 = ( \mbf b^\top \mbf r) \, ( \mbf r^\top \mbf b) 
	= \mbf b^\top ( \mbf r \, \mbf r^\top ) \mbf b \quad \mbox{and so} \quad
	\E ( \mbf r^\top \mbf b) ^2 = \mbf b^\top \mbf R \mbf b 
\end{equation}
where $ \mbf R$ denotes the symmetric square matrix $ \E [ \mbf r \mbf r^\top ]$ 
of second moments of returns, which we also assume exists.
The matrix $ \mbf R$ is positive definite under the assumption 
that the second moment $ \E ( \mbf r^\top \mbf b) ^2 $ of the return 
to any portfolio $ \mbf b \ne \mbf 0$ is always positive.

Substituting from (\ref{eq:corr}) and (\ref{eq:rbsq}) in (\ref{eq:expsq}) gives
\begin{eqnarray} 
	 \E ( \mbf p_2^\top \mbf a_2 - m_2 - \mbf r^\top \mbf b) ^2
 &=& \E ( \mbf p_2^\top \mbf a_2 - m_2 ) ^2
	- 2 ( \mbf p_2^\top \bar {\mbf a}_2 - \bar m_2 ) \bar { \mbf r}^\top \mbf b
	+ \mbf b^\top \mbf R \mbf b \nonumber \\
 &=& c + ( \mbf b^* - \mbf b)^\top \mbf R ( \mbf b^* - \mbf b) \label{eq:expects}
\end{eqnarray}
where \( { \mbf b^* }^\top \mbf R 
	= ( \mbf p_2^\top \bar {\mbf a}_2 - \bar m_2 ) \bar { \mbf r}^\top \),
implying that
 \( \mbf b^* = \mbf R^{-1} \bar { \mbf r}( \mbf p_2^\top \bar {\mbf a}_2 - \bar m_2 ) \),
and also
\begin{equation} \label{eq:valueOfc}
	c = \E ( \mbf p_2^\top \mbf a_2 - m_2 ) ^2 - { \mbf b^* }^\top \mbf R { \mbf b^* }
	= \E ( \mbf p_2^\top \mbf a_2 - m_2 ) ^2 
	- ( \mbf p_2^\top \bar { \mbf a}_2 - \bar m_2 ) ^2 
	\bar { \mbf r}^\top \mbf R^{-1} \bar { \mbf r}
\end{equation}
Finally, therefore, after ignoring an irrelevant additive constant,
the consumer's first-period maximand can be written as the quadratic form
\begin{equation} \label{eq:vxb}
	v( \mbf x_1, \mbf b) =
	- \tfrac 12 ( \mbf x_1 - \mbf a_1 )^\top \mbf Q_1 ( \mbf x_1 - \mbf a_1 ) 
 	- \tfrac 12 ( \mbf b^* - \mbf b)^\top \mbf S ( \mbf b^* - \mbf b)
\end{equation}
where $ \mbf S := \mbf R / \mbf p_2^\top \mbf Q_2 ^{-1} \mbf p_2 $.			

\subsection{The First-Period Optimization Problem}

The consumer's first-period optimization is therefore to maximise the function (\ref{eq:vxb})
w.r.t.\ $ \mbf x_1 $ and $ \mbf b$, 
subject to the budget constraint $ \mbf p_1^\top \mbf x_1 + \mbf q^\top \mbf b = m_1 $.
We solve this constrained maximization problem by introducing the Lagrangian
\begin{multline} \mathcal L_{ \lambda _1 }( \mbf x_1, \mbf b)
 = - \tfrac 12 ( \mbf x_1 - \mbf a_1 )^\top \mbf Q_1 ( \mbf x_1 - \mbf a_1 ) 
 - \tfrac 12 ( \mbf b^* - \mbf b)^\top \mbf S ( \mbf b^* - \mbf b) \\
 - \lambda _1 ( \mbf p_1^\top \mbf x_1 + \mbf q^\top \mbf b - m_1 ) 
\end{multline}
which is concave as a function of $( \mbf x_1, \mbf b)$, 
so is maximized w.r.t.\ $( \mbf x_1, \mbf b)$ when the two first-order conditions
\begin{equation} \label{eq:2FOCs}
\begin{array} {rcccl}
 \mbf 0 & = & \mathcal L'_{ \lambda _1, \mbf x_1 } 
 & = & - ( \mbf x_1 - \mbf a_1 )^\top \mbf Q_1 - \lambda _1 \mbf p_1^\top \\
 \quad \mbox{and} \quad \mbf 0 & = & \mathcal L'_{ \lambda _1, \mbf b} 
 & = & ( \mbf b^* - \mbf b)^\top \mbf S - \lambda _1 \mbf q^\top  
\end{array}
\end{equation}
are both satisfied.
Because both $ \mbf Q_1 $ and $ \mbf S$ are positive definite and so invertible, 
these first-order conditions are equivalent to
\begin{equation} \label{eq:solnFOCs}
 ( \mbf x_1 - \mbf a_1 )^\top = - \lambda _1 \mbf p_1^\top \mbf Q_1 ^{-1} 
 \quad \mbox{and} \quad
 ( \mbf b^* - \mbf b)^\top = \lambda _1 \mbf q^\top \mbf S^{-1} 
\end{equation} 
or, after transposing and rearranging, to
\begin{equation} \label{eq:solnA}
	\mbf x_1 = \mbf a_1 - \lambda _1 \mbf Q_1 ^{-1} \mbf p_1 \quad \mbox{and} \quad
 	\mbf b = \mbf b^* - \lambda _1 \mbf S^{-1}\mbf q 
\end{equation}
Substituting these into the budget equation gives
\begin{equation} \label{eq:solnB}
	\mbf p_1^\top \mbf x_1 + \mbf q^\top \mbf b 
	= \mbf p_1^\top ( \mbf a_1 - \lambda _1 \mbf Q_1 ^{-1} \mbf p_1 )
	+ \mbf q^\top ( \mbf b^* - \lambda _1 \mbf S^{-1} \mbf q)	= m_1 
\end{equation}
implying that 
\begin{equation} \label{eq:solnC}
	\lambda _1 = \frac { \mbf p_1^\top \mbf a_1 + \mbf q^\top \mbf b^* - m_1 }
	{ \mbf p_1^\top \mbf Q_1 ^{-1} \mbf p_1 + \mbf q^\top \mbf S^{-1} \mbf q} \end{equation}
Note that this is well defined because $ \mbf p_1 \ne \mbf 0$ and $ \mbf q \ne \mbf 0$,
whereas both symmetric matrices $ \mbf Q_1 $ and $ \mbf S$ are positive definite,
and so invertible with inverses that are positive definite.%
\footnote{A standard result in matrix theory
is that any symmetric $n \times n$ matrix $ \mbf A$ can be diagonalized,
in the sense that there exists an orthogonal matrix $ \mbf E$ 
(meaning that $ \mbf E^{-1} = \mbf E^\top $) 
and a diagonal matrix $ \mbf D$ such that $ \mbf {EAE^\top = D}$ and so $ \mbf {A = E^\top DE}$.
Then it is easy to see that the following four conditions are all logically equivalent:
(i) $ \mbf A$ is positive definite; (ii) all diagonal elements of $ \mbf D$ are positive;
(iii) $ \mbf D^{-1} $ exists and all its diagonal elements are positive;
(iv) $ \mbf A^{-1} = \mbf {E^\top D^{-1} E} $ exists and is positive definite.
}
Finally, we can use \eqref{eq:solnA} and \eqref{eq:solnC}
in order to determine the optimal commodity and asset demand vectors, which are
\begin{align} \label{eq:optimalDemands}
	\mbf x^*_1 &= \mbf a_1 - \frac { \mbf p_1^\top \mbf a_1 + \mbf q^\top \mbf b^* - m_1 }
	{ \mbf p_1^\top \mbf Q_1 ^{-1} \mbf p_1 
	+ \mbf q^\top \mbf S^{-1} \mbf q} \mbf Q_1 ^{-1} \mbf p_1 \\
 \mbox{and} \quad
 \mbf b &= \mbf b^* - \frac { \mbf p_1^\top \mbf a_1 + \mbf q^\top \mbf b^* - m_1 }
	{ \mbf p_1^\top \mbf Q_1 ^{-1} \mbf p_1 + \mbf q^\top \mbf S^{-1} \mbf q} \mbf S^{-1} \mbf q
\end{align}
Of course, for this solution to be economically sensible, 
we should require that $ \lambda _1 \ge 0$,
or equivalently, that 
\begin{equation} \label{eq:solnD}
	\mbf p_1^\top \mbf a_1 + \mbf q^\top \mbf b^* = \mbf p_1^\top \mbf a_1 
 	+ \mbf q^\top \mbf R^{-1} \bar { \mbf r} \, ( \mbf p_2^\top \bar {\mbf a}_2 - \bar m_2 ) 
	\ge m_1 
\end{equation} 
Furthermore, for the second-period solution we found previously to be economically sensible, 
we should require also that $ \mbf p_2^\top \mbf a_2 \ge m_2 + \mbf r^\top \mbf b$.
Because the two random variables $ \mbf p_2^\top \mbf a_2 - m_2 $
and $ \mbf r^\top \mbf b$ are independent, 
this requirement implies that there must be a real number $ \alpha $ for which, 
given the optimal choice of $ \mbf b$, one has
\begin{equation} \label{eq:solnE}
	\mbf p_2^\top \mbf a_2 - m_2 \ge \alpha \ge \mbf r^\top \mbf b 
\end{equation} 
for almost all possible values 
of the random pair $( \mbf p_2^\top \mbf a_2 - m_2, \mbf r^\top \mbf b) \in \R^2 $.

\subsection{An Enlivened Decision Problem} \label{ss:enlivenPortfolio}

To enliven this linear--quadratic decision model, we consider the possibility 
that unforeseeable changes occur after the pair $( \mbf x_1, \mbf b)$ 
has already been chosen in period~1.
In general, there could be a new second period objective 
\begin{equation} \label{eq:newProb}
	- \tfrac 12 ( \mbf x^+_2 - \mbf a^+_2 )^\top \mbf Q^+_2 ( \mbf x^+_2 - \mbf a^+_2 ) 
\end{equation}
in which the dimension of the vectors $ \mbf x^+_2 $, $ \mbf a^+_2 $ 
and the corresponding dimension 
of the positive definite square matrix $ \mbf Q^+_2 $ may have increased,
perhaps because of new commodities.
Of course, the second-period budget constraint must also change; we write it as
\begin{equation} \label{eq:newBudgConstraint}
	 \mbf p^+_2 {}^\top \, \mbf x^+_2 \le \mbf r^\top \, \mbf b + m_2 
\end{equation}
with the same asset vector $ \mbf b$ as before, since that is already determined
by the consumer's decisions in period~1.
The joint distribution of the triple $( \mbf a^+_2, m_2, \mbf r)$ may also change,
as indeed it must if the dimension of $ \mbf a^+_2 $ exceeds that of $ \mbf a_2 $.

If these changes could be known in advance, 
then in period 1 the consumer would face the problem of maximizing, 
instead of the quadratic evaluation function $ v( \mbf x_1, \mbf b) $ defined by (\ref{eq:vxb}), 
a revised quadratic objective function
\begin{equation} \label{eq:vxbplus}
		v^+ ( \mbf x_1, \mbf b) =
	- \tfrac 12 ( \mbf x_1 - \mbf a_1 )^\top \mbf Q_1 ( \mbf x_1 - \mbf a_1 ) 
 	- \tfrac 12 ( \mbf b^{+*} - \mbf b)^\top \mbf S^+ ( \mbf b^{+*} - \mbf b)
\end{equation}  
of the same choice variables $ \mbf x_1 $ and $ \mbf b$, subject to the same 
first-period budget constraint $ \mbf p_1^\top \mbf x_1 + \mbf q^\top \mbf b = m_1 $
as in \eqref{eq:budgEqa}.
What has changed, however, 
are the vector parameter $ \mbf b^{+*} $ and matrix parameter $ \mbf S^+ $ 
which appear in the last term of \eqref{eq:vxbplus},
whose changes are now entirely unpredictable.
Enlivenment requires recognizing that these parameters 
must be treated as themselves uncertain.
A Bayesian rational consumer who remains convinced 
that some quadratic model is still appropriate 
will, by definition, hold some subjective probability beliefs concerning
the unpredictable pair $( \widetilde { \mbf b}^*, \widetilde { \mbf S} )$ of parameters 
that characterize each member of the parametric family of quadratic evaluation functions
\begin{equation} \label{eq:tildev}
	\widetilde v( \mbf x_1, \mbf b; \widetilde { \mbf b}^*, \widetilde { \mbf S} ) 
	\equiv - \tfrac 12 ( \mbf x_1 - \mbf a_1 )^\top \mbf Q_1 ( \mbf x_1 - \mbf a_1 ) 
 	- \tfrac 12 ( \widetilde { \mbf b}^* - \mbf b)^\top \widetilde { \mbf S}
	( \widetilde { \mbf b}^* - \mbf b)
\end{equation}
Rationality, in the sense of subjective expected utility maximization, 
requires optimal policy in period~1 to maximize the expected value 
 \( \widehat \E [ \widetilde v( \mbf x_1, \mbf b; \widetilde { \mbf b}^*, 
 \widetilde { \mbf S} )] \)
of the function (\ref{eq:tildev}) w.r.t.\ probabilistic beliefs 
concerning the parameter pair $( \widetilde { \mbf b}^*, \widetilde { \mbf S} )$.
Such an expectation, however, after ignoring an irrelevant additive constant, 
can be expressed in the convenient form
\begin{equation} \label{eq:tildevxb}
	\widehat \E 
	[ \widetilde v( \mbf x_1, \mbf b; \widetilde { \mbf b}^*, \widetilde { \mbf S} )]
	\equiv - \tfrac 12 ( \mbf x_1 - \mbf a_1 )^\top \mbf Q_1 ( \mbf x_1 - \mbf a_1 ) 
 	- \tfrac 12 ( \widehat { \mbf b}^* - \mbf b)^\top 
	\widehat { \mbf S} ( \widehat { \mbf b}^* - \mbf b)
\end{equation}
This involves the appropriate subjective expected value 
 \( \widehat { \mbf S} := \widehat \E [ \widetilde { \mbf S} ] \) 
of the random matrix $ \widetilde { \mbf S} $.
Note that the matrix $ \widehat { \mbf S} $ is positive definite, and so invertible,
as the expected value of the random positive definite matrix $ \widetilde { \mbf S}$.%
\footnote{To show this, note that if the random symmetric matrix $ \widetilde { \mbf S} $
is almost surely positive definite, then for all $ \mbf u \in \R^n \setminus \{ \mbf 0 \}$
one has $ \mbf u^\top ( \widehat \E \widetilde { \mbf S} ) \mbf u 
= \widehat \E ( \mbf u^\top \widetilde { \mbf S} \mbf u) > 0$. 
}
This allows the vector $ \widehat { \mbf b}^* $ to be chosen uniquely 
so that it satisfies the first-order condition
 $ \widehat { \mbf S} \, \widehat { \mbf b}^* 
 = \widehat \E [ \widetilde { \mbf S} \, \widetilde { \mbf b}^* ]$.
This implies that there is a unique optimal vector $ \widehat { \mbf b}^* $ given by
\begin{equation} \label{eq:bstar}
	\widehat { \mbf b}^* = \widehat { \mbf S}^{-1} \, 
	\widehat \E [ \widetilde { \mbf S} \, \widetilde { \mbf b}^* ] 
	= ( \widehat \E [ \widetilde { \mbf S} ])^{-1} \, 
	\widehat \E [ \widetilde { \mbf S} \, \widetilde { \mbf b}^* ]
\end{equation}
Of course, equation \eqref{eq:bstar} typically implies 
that $ \widehat { \mbf b}^* \ne \widehat \E [ \widetilde { \mbf b}^* ]$ except,
for example, in the obvious special case 
when the pair $( \widetilde { \mbf b}^*, \widetilde { \mbf S} )$ of random parameters 
is uncorrelated, so 
 \( \widehat \E [ \widetilde { \mbf S} \, \widetilde { \mbf b}^* ]
  = \widehat \E [ \widetilde { \mbf S} ] \, \widehat \E [ \widetilde { \mbf b}^* ] \).

\section{Computer Chess} \label{s:chess}
\subsection{Simplified Chess}

The second of our two extended examples concerns the decision problem faced by a chess player 
who has to choose a move when confronted by a known position denoted by $ n_0 $.
To specify this position requires saying whose turn it is to move, and what piece, if any, 
occupies each of the 64 squares on the board.%
\footnote{Actually, even in a simplified version of chess
--- without either clocks that are used to enforce limits on each player's total thinking time, 
or drawing rules that go beyond stalemate, threefold repetition, or perpetual check
--- the rules of chess specify that: \\
(i) castling is disallowed if either the king or relevant rook has ever been moved previously; \\
(ii) a pawn can capture an opposition pawn \textit{en passant}, 
but only immediately after the pawn that is about to be captured 
has advanced two squares from its initial position.
So there are many chess positions whose full description requires significantly more information.}
Then let $ N_1 := N_{+1} ( n_0 )$ denote the set consisting of all those positions 
that can be reached by a move which is legal in position $ n_0 $.

Recall that, in the game of Chess, a player's King is in check 
just in case it is attacked by an opponent's piece,
in the sense that, in the absence of an intervening move, that piece could capture the King.
A player's move is legal only if it does not leave that player's King in check.
If the player whose turn it is to move has no legal move, then:
(i) either that player's King in check, 
in which case that player has been checkmated and loses the game;
(ii) or that player's King is not in check, 
in which case there is a stalemate and the game is a draw.

Following the famous result of Zermelo (1913), 
as well as von Neumann's (1928) pioneering analysis of maximin or minimax strategies 
in two-person ``zero-sum'' games of perfect information,
given best play by both the White and Black players, 
there is an objective \emph{result function}
\begin{equation} \label{eq:resultFn}
	N_1 \owns n_1 \mapsto r^+ ( n_1 ) \in \{W, D, L\}
\end{equation}
This function maps each possible position $ n_1 \in N_1 $ 
to a determinate \emph{result} $ r^+ ( n_1 ) \in \{W, D, L\} $ of the game that,  
for the player who is about to move, is either a win ($W$), or a draw ($D$), or a loss ($L$).
This result can be converted into a payoff using a scoring rule such as 1 for a win for White, 
or $-1$ for a win for Black, but 0 for a draw.
Then, given a continuation subgame of Chess that starts from the position $n$, 
the result of best play by both players in that subgame will be given 
by an objective \emph{evaluation function}
\begin{equation} \label{eq:normPayoffFn}
	N_1 \owns n_1 \mapsto v( n_1 ) \in \{ 1, 0, -1 \}  
\end{equation}

For the player whose turn it is to move at $ n_0 $, 
a move from $ n_0 $ to $ n_1 $ is optimal if and only if:
\begin{enumerate}
	\item $ n_1 $ maximizes the evaluation function $ v( n_1 )$ 
	in case it is White's turn to move at $ n_0 $;
	\item $ n_1 $ minimizes the evaluation function $ v( n_1 )$ 
	in case it is Black's turn to move at $ n_0 $.
\end{enumerate}

The objective normalized valuation function in \eqref{eq:normPayoffFn} can only be computed, 
however, for a few relatively simple positions where:
\begin{itemize}
	\item either it can be proved that, in a small number of moves, 
	one side can force a win due to checkmate, 
	or else, should they wish, a draw due to either
	(i) stalemate; (ii) a threefold repetition of the position; (iii) perpetual check;
	\item or alternatively, there are no more than 7 pieces on the board, 
	including both Kings, in which case the Lomonosov ``endgame tablebase'' software 
	cited in footnote 7 of Section \ref{ss:ubvbr} will specify 
	what is the result of the game if both players follow maximin strategies.
\end{itemize}
Thus, in choosing what move to make at $ n_0 $, 
and so what should be the next position $ n_1 \in N_1 $ on the board,
a player is typically forced to come up with subjective beliefs regarding the payoff function.
These beliefs can be guided by looking ahead a few moves.
But unless one can calculate with certainty a way to force 
a simple position whose evaluation is definitely known,
ultimately one has to assign such evaluations to many such positions a few moves ahead.
In this way, one constructs a subjective \emph{evaluation function} mapping chess positions 
into subjectively expected payoffs.
Computer chess programs for doing this involve algorithms that are good, even superhuman, 
but are still necessarily imperfect.
Currently some of the most effective software uses an algorithm 
based on \emph{Monte Carlo tree search} (MCTS), 
which is further discussed in Section \ref{ss:stockfishChess}
--- see Browne \textit{et al.}\ (2012) for a general survey 
that has been widely cited in the computer science literature.
Applied to Chess, in order to evaluate a given position $n$, 
MCTS considers many simulated continuation subgames that all start in position $n$, 
but then introduces a little carefully controlled randomness 
into the routine for choosing each ensuing move.
Then the final evaluation of any position $n$ is the average score 
over all the simulated games that start in position $n$.

\subsection{Real Chess}

Real chess is considerably more complicated. 
For one thing, a player about to move can claim a draw by demonstrating that the next move
can be chosen either to repeat the same position a third time, or so that both players 
will have made at least 50 moves without either a piece being captured or a pawn being moved.
Also, the game usually ends with either: (i) one player who is losing choosing to resign;
or (ii) with the two players agreeing to a draw 
when they both judge that they have an insufficient chance of winning.

Finally, there are time limits monitored by a chess clock, or actually a coupled pair of clocks,
one for each player, which displays how much remaining total time that player has available
before the next time control.
Whenever either player has just made a move, 
they can press a lever that simultaneously stops their own clock and starts the opponent's.
These additional considerations make the description of any chess position $n$ rather more complicated,
since it must include, for instance, 
how much more time each player can use before they would lose on time.

\subsection{Human Failure in a Bounded Model} \label{ss:chessBlunder}

Human chess experts exercise their skill by focusing attention 
on only a small number of plausible moves in each position.
Given any legal chess position $ n_0 $, consider the set $ N_1 := N_{+1} ( n_0 )$ 
of all possible positions $ n_1 $ that can result
after a legal move to $ n_1 $ is made from the position $ n_0 $.
Chess experts discern that many members $ n_1 $ of $ N_1 $, though allowable,
are too inferior to deserve much, if any, consideration.
Of course, human chess experts are also very good at judging the value of any position $ n_1 $ 
that they might think of moving to.
In this sense, they have good bounded models.

But, being merely human, even the very best players' models and evaluations 
of different positions may sometimes be grossly deficient. 
Witness how in 2006 Vladimir Kramnik, then the world champion, 
committed the ``blunder of the century'' by overlooking a checkmate in one move,
which led to an immediate loss.
This blunder was during the second game of a match of six games 
played against the computer program Deep Fritz.%
\footnote{%
See \url{https://en.chessbase.com/post/how-could-kramnik-overlook-the-mate-}.
Please note that the terminal hyphen is a key part of the address,
but there should be no full stop at the end.
}

\begin{figure} [hbt]
\begin{center}
\includegraphics[width=0.4\textwidth]{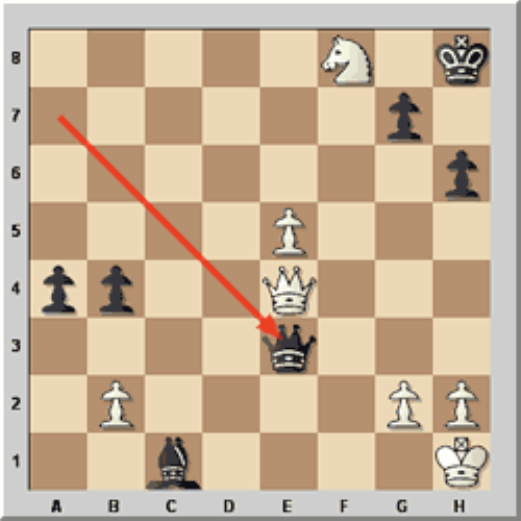}
\end{center}
\vspace{-2ex}
\caption{Deep Fritz v.\ Kramnik, Game 2} \label{fig:chess}
\end{figure}

In this game, Deep Fritz was playing with the White pieces.
Its last move before the position shown in Figure \ref{fig:chess} was its 34th.
The move was 34.\ Ne6$\times$f8.
This notation signifies that White's knight, which had been on square e6,
was used to take the Black piece, actually a rook, which had been on square f8.
In response, Kramnik (as Black) blundered horribly 
by playing the queen move 34 \ldots Qa7--e3,
as indicated by the arrow in Figure \ref{fig:chess}, 
thereby reaching the position shown in that Figure.
Whereupon the computer program Deep Fritz promptly indicated that its next move,
the queen move 35.\ Qe4--h7, would win at once by giving checkmate for White.

There must be thousands if not millions of chess players not nearly as strong as Kramnik 
who, if they were to be shown the position immediately before the move 34 \ldots Qa7--e3, 
would certainly notice that White was threatening to make 
the move 35.\ Qe4--h7 checkmate.
So how did Kramnik overlook it?

A good clue lies in the following observation that Kramnik himself offered 
during the press conference that was held minutes after the game ended:
\begin{quote}
``It was actually not only about the last move. \ldots \ 
I calculated the line many, many times, rechecking myself. 
I already calculated this line when I played 29 \ldots Qa7, 
and after each move I was recalculating, again, and again, 
and finally I blundered mate in one. 
Actually it was the first time that it happened to me, 
and I cannot really find any explanation. 
I was not feeling tired, I think I was calculating well during the whole game \ldots \ 
It's just very strange, I cannot explain it.''
\end{quote}
One way to interpret this is that Kramnik as Black, already when choosing his 29th move, 
had been planning to make what turned out subsequently 
to be the disastrous move 34 \ldots Qa7--e3 which is shown in Figure \ref{fig:chess}.
After all, if in Figure \ref{fig:chess} the Black king had been on square g8 rather than h8, 
then Black could have responded to 35.\ Qe4--h7 check with the move 35. \ldots K$\times$f8.
In the actual game, however, even at the last moment 
when he was about to make the move 34 \ldots Qa7--e3, 
he failed to modify his mental model of the game, as he should have,
by recognizing that this move allowed a mate in one move by his opponent.
In the end it was as if Kramnik had become so fixated 
on the earlier plan to choose 34 \ldots Qa7--e3 
that he failed to recognize the possibility of being checkmated 
immediately after making that fatal move.

Finally, although Kramnik claimed that he could not explain his blunder, 
in fact the quote above does offer a partial explanation.
What it misses is any explanation of why he became so fixated on his earlier flawed model.

\section{Bayesian Rationality in Decision Trees} \label{s:bayesRat}
\subsection{Roulette Lotteries} \label{ss:rouletteLotts}

Following the terminology of Anscombe and Aumann (1963),
given any non-empty set $Z$, let $ \Delta (Z) $ denote the set 
of all \emph{roulette lotteries} or \emph{simple probability measures}.
These take the form of functions $Z \owns z \mapsto \lambda (z) \in [0, 1]$
with a \emph{finite support} $ \supp \lambda \subseteq Z$ such that
\begin{equation} \label{eq:defSupp}
	\lambda (z) > 0 \Longleftrightarrow z \in \supp \lambda \quad \text{and} \quad
	\sum\nolimits _{z \in Z} \lambda (z) 
	= \sum\nolimits _{z \in \supp \lambda } \lambda (z) = 1
\end{equation}
Then, given any $z \in Z$, 
let $Z \owns z' \mapsto \delta _z (z') \in \Delta (Z) $ denote 
the unique \emph{degenerate} lottery that satisfies $ \delta _z (z) = 1$.
Also, whenever $Z$ is a finite set, let $ \Delta ^0 (Z) $ 
denote the set of \emph{fully supported} lotteries $ \lambda \in \Delta (Z) $
that satisfy $ \supp \lambda = Z$, 
or equivalently, $\lambda (z) > 0$ for all $z \in Z$.
Finally, given any $ \lambda, \mu \in \Delta (Z) $ 
and any scalar $ \alpha \in [0, 1]$,
let $ \nu := \alpha \, \lambda + (1 - \alpha ) \, \mu \in \Delta (Z) $ 
denote the \emph{lottery mixture} $Z \owns z \mapsto \nu (z) \in [0, 1] $ 
which, for all $z \in Z$, satisfies
\begin{equation} \label{eq:lottmix}
	\nu (z) = [ \alpha \, \lambda + (1 - \alpha ) \, \mu ] (z) 
	= \alpha \, \lambda (z) + (1 - \alpha ) \, \mu (z) 
\end{equation}

\subsection{Anscombe--Aumann (AA) Consequence Lotteries} \label{ss:horseLotts}

The hypothesis of Bayesian rationality, 
or equivalently of subjective expected utility maximization,
applies when there is a non-empty \emph{state space} $S$ 
of possible states of the world $s$ 
on which the subjective probability mapping 
 \( S \owns s \mapsto p(s) \in [0, 1] \) is defined, 
where $ \sum _{s \in S} p(s) = 1$.
Following Anscombe and Aumann (1963) once again, we assume that $S$ is finite.
Also, following their terminology 
which was described in Section \ref{ss:justBayes},
any random process for determining an uncertain state of the world 
will be described as a ``horse lottery''.

Bayesian rationality concerns preferences 
over Anscombe--Aumann consequence lotteries.
By definition, these may involve both risk, due to roulette lotteries, 
and uncertainty, due to horse lotteries.
Let $Y$ denote a non-empty consequence domain, 
and then let $ \Delta (Y) $ denote the domain of roulette lotteries over $Y$.

Next, given the finite set $S$ of states $s$ and the consequence domain $Y$, 
for each state $s \in S$, let $ Y_s $ be a copy of $Y$.%
\footnote{This is the case of a \emph{state-independent consequence domain},
which we assume in order to simplify notation.
The more general case of a \emph{state-dependent consequence domain} occurs 
when $ Y_s $ depends on $s$.
In this case, let $ Y^\cup := \cup _{s \in S} Y_s $ 
denote the \emph{union domain} 
of all consequences $y$ that are feasible in some state $s \in S$.
Then there may be a \emph{state-dependent utility function} 
 \( D \owns (s, y) \mapsto u(s, y) \to \R \)
defined on the domain $D := \{ (s, y) \in S \times Y^\cup \mid y \in Y_s \}$ 
of feasible state--consequence pairs. 
Such state-dependent utility functions have been studied 
in Dr\`eze (1962), Karni (1985), Schervish et al.\ (1990), 
Dr\`eze and Rustichini (2004), and Seidenfeld et al.\ (2010).
See Hammond (1998b, 1999, 2022) for a unified treatment which generalizes 
the special case when the consequence domain is state-independent, 
and which derives a state-independent utility function 
even in the general case when the consequence domain is state-dependent.
}
Then let 
\begin{equation} \label{eq:defLSY}
	L^S (Y) := \prod\nolimits _{s \in S} \Delta ( Y_s ) 
	= \{ \langle \lambda _s \rangle _{s \in S} 
	\mid \forall s \in S: \lambda _s \in \Delta ( Y_s ) \} 
\end{equation}
denote the space of \emph{Anscombe--Aumann lotteries}, or \emph{AA lotteries},
in the form of lists $ \langle \lambda _s \rangle _{s \in S} $ 
or mappings $S \owns s \mapsto \lambda _s \in \Delta (Y) $.
Each such mapping specifies a combination of, 
first, a horse lottery that determines a state $s \in S$,
followed second by a state-dependent roulette lottery $ \lambda _s $ 
that determines a consequence $y \in Y$.

\subsection{Choice from Pair Sets and Base Preferences} \label{ss:basePrefs}

Let $ \mathcal F_{ \setminus \emptyset } ( L^S (Y) )$ denote 
the family of non-empty finite subsets of the AA-lottery domain $ L^S (Y) $.
A \emph{choice function} on this lottery domain is a mapping 
\begin{equation} \label{eq:choiceFn}
	 \mathcal F_{ \setminus \emptyset } ( L^S (Y) ) \owns F \mapsto C(F) 
	 \in \mathcal F_{ \setminus \emptyset } ( L^S (Y) )
\end{equation}
that, for each non-empty \emph{feasible set} 
 \( F \in \mathcal F_{ \setminus \emptyset } ( L^S (Y) ) \), 
determines a non-empty \emph{choice set} 
 \( C(F) \in \mathcal F_{ \setminus \emptyset } ( L^S (Y) ) \) 
satisfying $ C(F) \subseteq F$.

Corresponding to any choice function $F \mapsto C(F) $ 
on $ \mathcal F_{ \setminus \emptyset } ( L^S (Y) )$, 
its values when $F$ is a \emph{pair set} with $ \#F = 2$
determine a strict preference relation $ \succ _C $,
a strict dispreference relation $ \prec _C $,
and an indifference relation $ \sim _C $.
These three binary relations are defined 
so that for each pair $ \lambda ^S, \mu ^S \in L^S (Y) $, one has
\begin{equation} \label{eq:basePref}
	\lambda ^S \left\{ 
	\begin{array} {c} \succ _C \\ \sim _C \\ \prec _C \end{array} \right \} 
	\mu ^S	\quad \text{according as} \quad 
	C( \{ \lambda ^S, \mu ^S \} ) = \left\{ \begin{array} {c} 
	\{ \lambda ^S \} \\ \{ \lambda ^S, \mu ^S \} \\ \{ \mu ^S \} 
	\end{array} \right\} 
\end{equation}
Underlying the choice function $F \mapsto C(F) $ 
specified by \eqref{eq:choiceFn},
there is a single corresponding binary weak preference relation $ \succsim _C $
on $ L^S (Y) $, called the \emph{base relation}.
For each pair $ \lambda ^S, \mu ^S \in L^S (Y) $, this base relation satisfies
\begin{equation} \label{eq:baseWPref}
	\lambda ^S \succsim _C \mu ^S 
	\Longleftrightarrow \lambda ^S \in C( \{ \lambda ^S, \mu ^S \} )
	\Longleftrightarrow \lambda ^S \succ _C \mu ^S 
		\ \text{or} \ \lambda ^S \sim _C \mu ^S
\end{equation}
Finally, we mention the corresponding 
weak dispreference relation $ \precsim _C $ defined on $ L^S (Y) $ so that 
\begin{equation} \label{eq:dispref}
	\lambda ^S \precsim _C \mu ^S \Longleftrightarrow 
	\mu ^S \in C( \{ \lambda ^S, \mu ^S \} ) 
	\Longleftrightarrow \lambda ^S \prec _C \mu ^S  
		\ \text{or} \ \lambda ^S \sim _C \mu ^S
\end{equation}
Evidently both the weak preference relation $ \succsim _C $ 
and the weak dispreference relation $ \precsim _C $ are \emph{complete}
in the sense that, for each pair $ \lambda ^S, \mu ^S \in L^S (Y) $, 
both the following must hold:
\begin{enumerate}
	\item either $ \lambda ^S \succsim _C \mu ^S $, 
	or $ \mu ^S \succsim _C \lambda ^S $, or both;
	\item either $ \lambda ^S \precsim _C \mu ^S $, 
	or $ \mu ^S \precsim _C \lambda ^S $, or both.
\end{enumerate}

\subsection{Bayesian Rationality and Expected Utility} \label{ss:bayesianEU}

Let $F \mapsto C(F) $ be any choice function 
satisfying both \eqref{eq:choiceFn} and $ C(F) \subseteq F$ 
for all $F \in \mathcal F_{ \setminus \emptyset } ( L^S (Y) )$ 
that corresponds to the base preference relation $ \succsim _C $ 
defined on the space $ L^S (Y) $ of AA lotteries $ \lambda ^S $ 
by \eqref{eq:basePref}.
Then the mapping $ L^S (Y) \owns \lambda ^S \mapsto U^S ( \lambda ^S ) \in \R$ 
is a \emph{utility function} which \emph{represents} 
the preference relation $ \succsim _C $ on $ L^S (Y) $ just in case, 
for all $ \lambda ^S, \mu ^S \in L^S (Y) $, one has
\begin{equation} \label{eq:utilRepr}
	\lambda ^S \succsim _C \mu ^S 
	\Longleftrightarrow U^S ( \lambda ^S ) \ge U^S ( \mu ^S )
\end{equation}

The following definition departs 
from what has become standard in decision theory 
by excluding zero subjective probabilities. 
As discussed in Section 5.4 of Hammond (1998b) as well as in Hammond (2022),
this restriction is imposed in order to avoid the difficulties that arise 
in continuation subtrees of a decision tree 
when zero probabilities are allowed. 

\begin{definition}
Let $S$ denote the fixed non-empty finite set of uncertain states of the world.
\begin{enumerate}
	\item An \emph{interior subjective probability mass function} 
	is a mapping $S \owns s \mapsto \mathbb P(s) \in (0, 1] $ 
	that satisfies $ \sum _{s \in S} \mathbb P(s) = 1$.
	\item The choice function $F \mapsto C(F) $ 
	on the domain $ \mathcal F_{ \setminus \emptyset } ( L^S (Y) )$,
	together with the associated base preference relation $ \succsim _C $ 
	on the space $ L^S (Y) $, are both \emph{Bayesian rational} just in case 
	there exist an interior subjective 
	probability mass function $S \owns s \mapsto \mathbb P(s) \in (0, 1] $, 
	as well as 
	a \emph{Bernoulli utility function} $Y \owns y \mapsto u(y) \to \R $,
	such that $ \succsim _C $ is represented, 
	in the sense that \eqref{eq:utilRepr} is satisfied, 
	by the \emph{von Neumann subjective expected utility function} defined 
	for all AA lotteries 
 \( \lambda ^S = \langle \lambda _s \rangle _{s \in S} \in L^S (Y) \) 
	by the double sum
\begin{equation} \label{eq:subjEU}
	U^S ( \lambda ^S ) = \sum\nolimits _{s \in S} 
	\mathbb P(s) \sum\nolimits _{y \in Y} \lambda _s (y) u(y) 
\end{equation}
\end{enumerate}
\end{definition}

\subsection{Normalized Utility} \label{ss:constructBUF}

Recall that, as explained in Section \ref{ss:rouletteLotts}, 
for each $y \in Y$, we use $ \delta _y $ to denote 
the unique \emph{degenerate} probability measure in $ \Delta (Y) $ 
that satisfies $ \delta _y ( \{y\}) = 1$.
To avoid trivialities, we assume that there exist 
at least three consequences $ \underline y, y^0, \overline y$ in the domain $Y$ 
such that, for the three corresponding degenerate lotteries 
 \( \delta _{ \overline y}, \delta _{ y^0 }, \delta _{ \underline y} \),
the base strict preference relation $ \succ $ 
satisfies the strict preference property
 \( \delta _{ \overline y} \succ \delta _{ y^0 } 
 \succ \delta _{ \underline y} \).

Consider any Bernoulli utility function $Y \owns y \mapsto u(y) \to \R $
and the associated preference relation $ \succsim $ on $ \Delta (Y) $ 
that together satisfy 
\begin{equation} \label{eq:expectBUF}
	\lambda \succsim \mu \Longleftrightarrow U( \lambda ^S ) \ge U( \mu ^S )
\end{equation}
for the \emph{von Neumann objective expected utility function} 
on $ \Delta (Y) $ defined by 
\begin{equation} \label{eq:objEU}
	U( \lambda ) = \sum\nolimits _{y \in Y} \lambda (y) u(y) 
\end{equation}
As explained in Hammond (1998a, Section 2.3), assuming Bayesian rationality,
the ratio \( \dfrac { u( y^0 ) - u( \underline y) }
 { u( \overline y) - u( \underline y) } \) of utility differences equals, 
in economists' terminology, the constant marginal rate of substitution 
along an indifference curve between shifts in probability:
(i) from consequence $ \underline y$ to $ y^0 $;
(ii) from consequence $ \underline y$ to $ \overline y$.
This leads us to say 
that two Bernoulli utility functions $y \mapsto u(y) $ 
and $y \mapsto \tilde u(y) $ are \emph{equivalent} just in case, 
for every triple $ \underline y, y^0, \overline y$ of consequences in $Y$ 
satisfying 
\( \delta _{ \overline y} \succ \delta _{ y^0 } \succ \delta _{ \underline y} \) 
and so $ u( \overline y) > u( y^0 ) > u( \underline y)$,
the corresponding ratios of utility differences satisfy
\begin{equation} \label{eq:equalRatios}
	\frac { u( y^0 ) - u( \underline y) }
	{ u( \overline y) - u( \underline y) } 
	= \frac { \tilde u( y^0 ) - \tilde u( \underline y) }
	{ \tilde u( \overline y) - \tilde u( \underline y) }
\end{equation}
But \eqref{eq:equalRatios} holds 
for every triple $ \underline y, y^0, \overline y$ 
satisfying $ u( \overline y) > u( y^0 ) > u( \underline y)$ 
if and only if there exist an additive constant $ \alpha \in \R$ 
and a positive multiplicative constant $ \rho \in \R$ such that, 
for all $y \in Y$, one has 
\begin{equation} \label{eq:affineEq}
	\tilde u(y) = \alpha + \rho u(y) 
\end{equation}

Now, given any pair $ \underline u, \bar u$ of real numbers 
with $ \bar u > \underline u$
as well as any Bernoulli utility function $Y \owns y \mapsto u(y) \to \R $,
there exist two unique constants $ \alpha $ and $ \rho > 0$ 
such that the transformed utility function defined by \eqref{eq:affineEq} 
is a \emph{normalized} utility function 
that satisfies $ \tilde u( \underline y) = \underline u$
and $ \tilde u( \overline y) = \overline u$.
Indeed the two constants we need are given by
\begin{equation} \label{eq:normConsts}
	\rho 
	= \frac { \overline u - \underline u}{u( \overline y) - u( \underline y)} 
 \quad \text{and then} \quad 
 	\alpha 
	= \overline u - \rho u( \overline y) = \underline u - \rho u( \underline y)
\end{equation}
From now on let $u$ denote 
the unique Bernoulli utility function $Y \owns y \mapsto u(y) \to \R $
whose expected values defined by \eqref{eq:objEU} satisfy \eqref{eq:expectBUF},
and which has been normalized to satisfy  
\begin{equation} \label{eq:normBUF}
	u( \underline y) = \underline u \quad \text{and} 
	\quad u( \overline y) = \overline u
\end{equation}

\subsection{Finite Decision Trees and Their Continuations} \label{ss:decTree}

In mathematical terminology, a \emph{finite directed graph} $(N, E)$ combines:
\begin{enumerate}
	\item a non-empty finite set $N$ of vertices or \emph{nodes} $n$;
	\item a specified subset $E \subseteq N \times N$ 
	of \emph{directed edges} consisting of ordered pairs $e = (n, n')$,
	alternatively denoted by $n \to n'$, with $n \ne n'$. 
\end{enumerate}

The sequence $( n_1, n_2, \ldots, n_\ell )$ of~$ \ell $ nodes in $N$
is a \emph{path} of length $ \ell \in \N$ in the directed graph $(N, E)$
just in case $ n_k \to n_{k + 1} $ is a directed edge in $E$ 
for $k = 1, 2, \ldots, \ell - 1$.

The graph $(N, E)$ is a \emph{rooted directed tree} just in case 
there is a unique \emph{initial node} $ n_0 \in N$ 
(or root, or seed, or entry point)
such that, for every other node $n \in N \setminus \{ n_0 \}$ of the graph, 
there is a unique path $( n_0, n_1, n_2, \ldots, n)$ 
which starts at node $ n_0 $ and ends at node $n$.

\begin{definition}
	Let $T = (N, E)$ denote any finite rooted directed tree.
\begin{enumerate}
	\item Given any pair of nodes $n, n' \in N$, 
	say that $n' \in N$ (ultimately) \emph{succeeds} $n$ just in case 
	there is a path $( n_1, n_2, \ldots, n_\ell )$ of nodes in $N$ 
	of length $ \ell $ which joins $ n_1 = n$ to $ n_\ell = n'$.
	Let $ N_{> n} $ denote the set of all nodes $n' \in N$ that succeed $n$.
	\item For each $n \in N$, define 
	the \emph{continuation subtree} $ T_{\ge n} = ( N_{\ge n}, E_{\ge n} )$ 
	whose initial node is $n$ as the unique tree in which:
	\begin{itemize}
		\item $ N_{\ge n} = \{n\} \cup  N_{> n} $;
		\item $ E_{\ge n} $ is the restriction 
		to $E \cap ( N_{\ge n} \times N_{\ge n} )$ of edges in $E$. 
	\end{itemize}
	\item Say that any other node $n' \in N$ \emph{immediately succeeds} $n$ 
	just in case the ordered pair $(n, n')$ or $n \to n'$ 
	is a directed edge of $T$.
	\item Denote the set of all the immediate successors of $n$ by
\begin{equation} \label{eq:immedSuccs}
	N^{+1} _{ \ge n} := \{ n' \in N \mid (n, n') \in E \} 
\end{equation}
\end{enumerate}
\end{definition}

Recall from Section \ref{ss:justBayes} the distinction 
between roulette and horse lotteries due to Anscombe and Aumann (1963).
The following definition builds on those in Hammond (1988b; 1998a, b; 2022). 

\begin{definition} \label{def:treeDomain}
Given the non-empty consequence domain $Y$ 
and non-empty finite set $S$ of possible states of the world,
define $ \mathcal T^S (Y) $ as the collection 
of all finite \emph{decision trees} which combine 
a rooted directed tree $T = (N, E)$ with an \emph{event correspondence} 
 \( N \owns n \mapsto S_{ \ge n} \in 2^S \setminus \{ \emptyset \} \)
specifying what set $ S_{ \ge n} $ of states are possible 
in each continuation subtree $ T_{ \ge n} $.
Moreover, the set $N$ of nodes in the decision tree is partitioned 
into four pairwise disjoint subsets:
\begin{enumerate}
	\item the set $ N^d $ of \emph{decision nodes} $n$ at which:
	(i) at $n$ the agent must choose, 
	for some node $n'$ in the set $ N^{+1} _{ \ge n} $ 
	of immediate successors defined in \eqref{eq:immedSuccs}, 
	the directed edge $n \to n'$;
	(ii) $ S_{ \ge n'} = S_{ \ge n} $ for all $n' \in N^{+1} _{ \ge n} $;
	\item the set $ N^c $ of \emph{chance nodes} $n$ where:
	(i) a directed edge $n \to n'$ emanating from~$n$ is determined randomly 
	by a specified \emph{roulette lottery}
	 \( N^{+1} _{ \ge n} \owns n' \mapsto \pi ( n'| n) \in (0, 1] \);%
 \footnote{See Hammond (1988b) for an explanation of why, 
 if there is a chance node $n \in N^c $ and a node $n' \in N^{+1} _{ \ge n} $ 
 at which $ \pi ( n'| n) = 0$, 
 then both the consequentialist and the equivalent prerationality 
 justifications of Bayesian rationality 
 which were described in Section \ref{ss:justBayes} 
 imply that all consequence lotteries must be indifferent.}
	(ii) $ S_{ \ge n'} = S_{ \ge n} $ for all $n' \in N^{+1} _{ \ge n} $;
	\item the set $ N^e $ of \emph{event nodes} $n$ 
	where a directed edge $n \to n'$ emanating from~$n$ 
	is determined by a \emph{horse lottery} 
	whose outcome partitions the event $ S_{ \ge n} $
	into the collection $ \{ S_{ \ge n'} \mid n' \in N^{+1} _{ \ge n} \} $ 
	of non-empty pairwise disjoint sub-events $ S_{ \ge n'} $ which, 
	for each $ n' \in N^{+1} _{ \ge n} $, 
	consist of states $s \in S$ that can occur in the subtree $ T_{ \ge n'} $;
	\item the non-empty set $ N^t $ of \emph{terminal nodes} $n$ 
	at which $ N^{+1} _{ \ge n} = \emptyset $, so no edge emanates, 
	and which are each mapped
	to an \emph{Anscombe--Aumann consequence lottery} \( \gamma (n) 
 = \langle \gamma _s \rangle _{s \in S_{ \ge n} } \in L^{ S_{ \ge n} } (Y) \)
	whose outcomes $y$ belong to the specified consequence domain $Y$.
\end{enumerate}
\end{definition}

Say that a decision tree $T \in \mathcal T^S (Y) $ is:
\begin{itemize}
	\item \emph{deterministic} just in case $ N^c = N^e = \emptyset $;
	\item \emph{risky} just in case $ N^c \ne \emptyset $ 
	but $ N^e = \emptyset $;%
 \footnote{Raiffa (1968) focused on risky decision trees 
 with \emph{pecuniary consequences} 
 in the form of payoffs measured in dollars.} 
	\item a \emph{Savage tree} 
	just in case $ N^e \ne \emptyset $ but $ N^c = \emptyset $;
	\item an \emph{Anscombe--Aumann tree} 
	just in case $ N^c \ne \emptyset $ and $ N^e \ne \emptyset $.
\end{itemize}
	
\subsection{Evaluations of Continuation Subtrees} \label{ss:valueSubtrees}

Consider the orthodox ``unenlivened'' decision model which is represented 
by any finite decision tree $T$ in the domain $ \mathcal T^S (Y) $ of trees 
with non-empty finite state space $S$ and AA lottery consequences 
in the domain $ \cup _{S' \in 2^S \setminus \{ \emptyset \} } L^{ S' } (Y) $.
Then the justifications of Bayesian rationality 
which were described in Section \ref{ss:justBayes} 
imply that, as usual in dynamic programming,
one can use a recursive procedure that works backwards within each decision tree
in order to calculate the \emph{evaluation} $ v( T_{ \ge n} )$,
which is the subjectively expected continuation value 
of reaching the initial node $n \in N$ 
of the continuation subtree $ T_{ \ge n} $.
The details of this backward recursive procedure 
are described in the remainder of this section.

In any finite decision tree $T$, the backward recursion starts 
at any terminal node $n \in N^t $.
As discussed in Section \ref{ss:decTree}, the specified consequence 
of reaching any terminal node $n \in N^t $ of tree $T$ 
is the AA consequence lottery \( \gamma (n) 
 = \langle \gamma _s \rangle _{s \in S_{ \ge n} } \in L^{ S_{ \ge n} } (Y) \).
Then the evaluation $ v( T_{ \ge n} )$ of the subtree $ T_{ \ge n} $, 
whose only node is the terminal node $n$, 
is the expected utility of the AA consequence lottery $ \gamma (n) $,
which is specified by \eqref{eq:subjEU} as the double sum
\begin{equation} \label{eq:termVal}
	v( T_{ \ge n} ) = U^{ S_{ \ge n} } ( \gamma (n) ) 
	= \sum\nolimits _{s \in S_{ \ge n}} \mathbb P(s) 
	\sum\nolimits _{y \in Y} \gamma _s (y) \, u(y) 
\end{equation} 

At any non-terminal node $n \in N \setminus N^t $, 
the evaluation $ v( T_{ \ge n} )$ of the continuation subtree $ T_{ \ge n} $ 
starting at the initial node~$n$
depends upon the set $ \{ v( T_{ \ge n'} ) \mid n' \in N^{+1} _{ \ge n} \}$ 
of evaluations of all the continuation subtrees 
starting at a node $n' \in N^{+1} _{ \ge n} $ which immediately succeeds $n$.
Specifying the correct formula for $ v( T_{ \ge n} )$ 
requires considering separately three cases, 
depending upon whether $n$ is a chance, event, or decision node.

In the first case when $n$ is a chance node 
whose immediate successors $n' \in N^{+1} _{ \ge n} $ 
occur with respective specified positive probabilities $ \pi ( n'| n) $, 
the relevant recursion takes the obvious form
\begin{equation} \label{eq:chanceVal}
	v( T_{ \ge n} ) 
	= \sum\nolimits _{n' \in N^{+1} _{ \ge n} } 
		\pi ( n'| n) \, v( T_{ \ge n'} )
\end{equation} 

The second case occurs when $n$ is an event node,
each of whose immediate successors $n' \in N^{+1} _{ \ge n} $ 
determines which is the relevant cell 
In this case the objectively specified probabilities $ \pi ( n'| n) $ 
of the partition $ \{ S_{ \ge n'} \mid n' \in N^{+1} _{ \ge n} \}$ 
of the event $ S_{ \ge n} $ into pairwise disjoint sets.
that appear in \eqref{eq:chanceVal} need to be replaced 
by subjective conditional probabilities $p ( n'| n) $
derived from the relevant subjective probabilities $ \mathbb P(s) $ 
for different states $s \in S_{ \ge n}$.
Because of our requirement that $ \mathbb P(s) > 0$ for all $s \in S$,
these conditional probabilities $p ( n'| n) $ are all well defined, 
and can be calculated as
\begin{equation} \label{eq:subjCondProb}
	p ( n'| n) = \sum\nolimits _{s \in S_{ \ge n'} } \mathbb P(s) \ 
	/ \sum\nolimits _{s \in S_{ \ge n} } \mathbb P(s)
\end{equation} 
So, when $n$ is an event node with subjective probabilities $p ( n'| n) $ 
given by \eqref{eq:subjCondProb}
rather than a chance node 
with hypothetical or objective probabilities $ \pi ( n'| n) $,
the previous formula \eqref{eq:chanceVal} is changed to
\begin{equation} \label{eq:eventVal}
	v( T_{ \ge n} ) 
	= \sum\nolimits _{n' \in N^{+1} _{ \ge n} } p( n'| n) \, v( T_{ \ge n'} )
\end{equation}

In the third and final case when $n$ is a decision node, we apply 
the standard \emph{optimality principle} of stochastic dynamic programming. 
This requires any current \emph{optimal decision} $ n^* \in N^{+1} _{ \ge n} $ 
to be the first step toward achieving the highest possible expected value 
resulting from an appropriate plan for all subsequent decisions.
Consider the induction hypothesis that, 
for each node $n' \in N^{+1} _{ \ge n} $, the value $ v( T_{ \ge n'} )$ 
is the maximum possible evaluation the agent can achieve 
by choosing an optimal decision at each decision node of $ T_{ \ge n'} $.
This is trivally true when $n'$ is a terminal node, 
so there is no decision to make at node $n'$.
If this hypothesis is true at each node $n' \in N^{+1} _{ \ge n} $,
then any optimal decision at node $n$ must be to move 
along an edge $n \to n^* $ to an immediately succeeding node $ n^* $ 
which maximizes the evaluation $v( T_{ \ge n'} )$ with respect to $n'$
subject to $n' \in N^{+1} _{ \ge n} $, 
where the set $ N^{+1} _{ \ge n} $ is finite. 
In other words, one must satisfy
\begin{equation} \label{eq:argmaxValue}
	n^* \in \argmax _{n' \in N^{+1} _{ \ge n}} v( T_{ \ge n'} )
\end{equation}
So the appropriate recursion when $n$ is a decision node is
\begin{equation} \label{eq:maxValue}
	v( T_{ \ge n} ) = v( T_{ \ge n^* } ) 
	= \max _{n' \in N^{+1} _{ \ge n}} \nolimits v( T_{ \ge n'} )
\end{equation}

Together, therefore, 
the four equations \eqref{eq:termVal}, \eqref{eq:chanceVal}, 
\eqref{eq:eventVal}, and \eqref{eq:maxValue} 
do indeed determine $v( T_{ \ge n} )$ 
by backward recurrence in the four different cases. 

\section{Truncated Decision Trees} \label{s:truncTrees}

\begin{quote}
	``The second factor which imposes a horizon 
	upon the imaginative creation of the future 
	is that uncertainty becomes more and more unbounded 
	by considerations of what is possible, 
	the more remote the date considered.''
	--- Shackle (1969, page 224)

\end{quote}

\subsection{Bounded Rationality as Bounded Modelling}

The conditionally expected enlivened evaluation 
associated with entering any continuation subtree $ T_{ \ge n}$ 
of a decision tree $T \in \mathcal T^S (Y) $ is $ v( T_{ \ge n} )$.
In principle, this can be calculated by following the procedure 
of backward recursion that was set out in Section \ref{ss:valueSubtrees}.
Carrying out all the required steps of the computation by backward recursion, 
however, obviously becomes much more challenging, 
if not practically impossible, as at least some 
of the continuation subtrees $ T_{ \ge n}$ become more complicated.

As discussed in Section \ref{ss:boundedModels}, 
these practical limitations on being able to calculate an optimal decision 
may force the decision maker to resort to some kind of bounded model
in the form of a bounded decision tree $ \bar T$.
Now, provided that every terminal node of the bounded tree 
is a terminal node of the original tree $T \in \mathcal T^S (Y) $,
one will have $ \bar T \in \mathcal T^S (Y) $.
Then the dynamic programming argument used in Section \ref{s:bayesRat}
can be applied to tree $ \bar T$ instead of~$T$.

In this section, we focus on the more challenging case involving bounded models 
that result when the original decision tree gets truncated 
by having some entire continuation subtrees removed,
including their terminal nodes with their consequences.
In the game of Chess, this is the kind of truncated tree that results 
when a player looks at most $k$ moves ahead, for some small $k \in \N$.

\subsection{Cuts and Truncation Nodes}

The following definition captures the idea that any truncation 
of a decision tree will occur at a truncation node.
This node has the property that all strictly succeeding nodes are removed,
so the truncation node becomes a terminal node.

\begin{definition}
Let $T$ be any decision tree in the domain $ \mathcal T^S (Y) $. 
The tree $ \hat T = ( \hat N, \hat E) $ is a \emph{truncation} of $T$ 
just in case there is a set $X$ of one or more \emph{cuts}
or non-terminal \emph{truncation nodes} $x \in N \setminus N^t $ such that:
\begin{enumerate}
	\item there is a pairwise disjoint collection $ \{ N_{> x} \mid x \in X \}$
	of removed sets $ N_{ >x } = N_{ \ge x } \setminus \{ x \}$ 
	of non-initial nodes $n \in N \setminus \{ n_0 \} $, 
	each associated with a continuation subtree $ T_{\ge x} $ 
	whose initial node is at a cut $x \in X$;
	\item the truncated set $ \hat N$ of remaining nodes
	is $N \setminus \cup _{x \in X} N_{ >x }$;
	\item the truncated set $ \hat E$ of remaining directed edges 
	is the restriction $( \hat N \times \hat N) \cap E$ 
	to $ \hat N \times \hat N$ of the original set $E \subset N \times N$ 
	of directed edges $n \to n'$ in $T$;
	\item each cut $x \in X$ is a terminal node $x \in \hat N^t $ 
	of the truncated tree $ \hat T$.
\end{enumerate}
\end{definition}

\subsection{Subjective Evaluations in Truncated Trees} \label{ss:truncatedEvals}

Given a finite decision tree $T = (N, E) $,
in order to make the truncation $ \hat T = ( \hat N, \hat E) $ of $T$ 
a decision tree, each terminal node $n \in \hat N^t $ of $ \hat T$ 
must be assigned a consequence $ \hat \gamma (n) $.
In case $n \in \hat N^t \cap N^t $, 
which is when the terminal node~$n$ of~$ \hat T$
happens also to be a terminal node of the untruncated tree $T$, 
the consequence $ \hat \gamma (n) $ of $n$ in~$ \hat T$
should be the same as its consequence in $T$, 
implying that $ \hat \gamma (n) = \gamma (n) $.

In case node $n$ is a cut $x \in \hat N^t \setminus N^t $, however,
no consequence $ \hat \gamma (n) = \hat \gamma (x) $ 
has yet been specified at node $n = x$,
which has become a terminal node of the truncated tree $ \hat T$ 
but not of the original untruncated tree $T$.
In the full tree $T$, with unbounded rationality,
an appropriate ideal evaluation $ v( T_{>x} )$ of this menu consequence,
as well as $ v( T_{ \ge n} )$ of the node $n = x \in \hat N^t \setminus N^t $,
is the expected utility level that is calculated by backward recursion, 
following the rules set out in Section \ref {ss:valueSubtrees}.
In a truncated tree $ \hat T$, however, the best the agent can do is to use 
a \emph{subjective evaluation} $ \hat \gamma (n) \in \R$.
This should be an estimate of what expected utility $ v( T_{ \ge n} )$ 
would have resulted from the calculation by backward recursion 
had the agent been able to carry this out 
in the full continuation subtree $ T_{ \ge n} $, without any truncation.%
\footnote{It may be better to replace the real-valued subjective evaluation 
attached to each truncation node with a subjective menu consequence 
of the kind considered in the forthcoming paper Hammond and Troccoli Moretti (2026)
--- see also Section \ref{ss:menuConseqs}.
We defer to later research further exploration of this idea.}

In the rest of the paper, to allow for truncated trees, we distinguish between:
\begin{enumerate}
	\item ``consequence'' terminal nodes $n$, 
	each of which has a specified Ans\-combe--Aumann 
	lottery consequence $ \gamma (n) \in L^{ S_{ \ge n}} (Y)$ attached;
	\item ``truncation'' nodes $x \in X$,  
	which each have a subjective evaluation $ \hat \gamma (x) \in \R $ 
	attached.
\end{enumerate}

\subsection{Monte Carlo Tree Search and Scenario Planning} \label{ss:stockfishChess}

\begin{quote}
	Long before computer games became popular recreations, 
	mathematicians viewed games	as models of decision making. 
	The general understanding of decisions, however, has been impeded 
	by the ambiguity of some of the basic components of game-tree search. 
	In particular, the static evaluation function, 
	or determination of a node's merit based on directly detectable features, 
	has never been adequately defined.
	The expected-outcome model proposes that the appropriate value 
	to assign a node is the expected value of a game's outcome 
	given random play from that node on. \\
	--- from the abstract to Bruce Abramson's (1987) Ph.D.\ dissertation, 
	eventually published as Abramson (1991).
\end{quote}

We turn next to a powerful kind of computer algorithm that has proved 
extremely fruitful in estimating normatively useful continuation values 
for at least some particular kinds of truncated game.
Indeed, exploiting Abramson's key idea of Monte Carlo tree search (MCTS)
described in the quotation above 
helped to inspire a generation of computer programs that:
\begin{enumerate}
	\item in the case of Chess, led to the Stockfish software engine 
	that would easily beat any human player 
	over any sufficiently long run of games;
	\item but in the case of Go, was unable to defeat the best human players.
\end{enumerate}

In decision trees, any relevant version of a procedure like MCTS 
evidently requires evaluating each continuation decision tree by estimating 
the normalized expected utility from a ``Monte Carlo'' sample of simulations 
in which the agent makes a suitably randomized choice 
at each decision node in that continuation.

This approach to valuing a continuation subtree 
which can never be completely modelled was the basis 
of the successful Deep Fritz and then Stockfish open source engines 
for computer chess.
Eventually, however, this generation of algorithms became superseded by Alpha\-Zero,
which combines MCTS and reinforcement learning 
with the kind of artificial neural network that has become 
a key part of what has come to be known as ``artificial intelligence''.
Algorithms like Alpha\-Zero have proved far better at playing Chess 
than programs of the Stockfish generation, 
while finally becoming able to beat the best human players at Go.
The key paper by Silver et al.\ (2018), however, 
reports that MCTS remains part of AlphaZero.%
\footnote{See also \url{https://en.wikipedia.org/wiki/AlphaZero} 
for more current details.}

A much more widely applicable procedure that has proved useful 
in some business applications involves ``scenario planning''.
This is somewhat similar to the rational shortlist method 
considered in Section \ref{ss:ratShortlist}.
It involves considering in detail a limited range 
of bounded incomplete scenarios which are selected 
in an attempt to pay appropriate attention to eventualities 
that are likely to have a significant bearing on the final choice of policy.
This is in contrast to what may be a standard view 
that the focus should be on the most likely eventualities.%
\footnote{See Jefferson (2012) for the point of view 
of the Royal Dutch/Shell Group's former Chief Economist, 
and his experience with scenario planning.
See also Jefferson (2014) and Derbyshire (2017) 
for discussion of the possible links between the two concepts 
of potential surprise due to Shackle (1953, 1969)  
and scenario planning as practised by Shell.}

\subsection{A Subjective Evaluation at Each Truncated Node} \label{ss:menuConseq}

To provide a basis for this kind of subjective estimate
considered in Section \ref{ss:truncatedEvals}, we extend 
the domain $ \mathcal T^S (Y) $ of finite decision trees with consequences 
in the original domain~$Y$ in order to assign to each cut node $x \in X$ 
a subjective evaluation $ \hat \gamma (x) \in \R $.
With this modification, the analysis of Bayesian rationality laid out 
in Section \ref{s:bayesRat} can be applied 
to the extended domain \( \mathcal T^S ( Y \cup \R ) \) of decision trees 
with consequences in the extended domain $Y \cup \R $.
This extended analysis shows that Bayesian rationality 
implies maximizing the expected value 
of a unique extended normalized Bernoulli utility function $ \hat u$
defined on the extended domain $Y \cup \R $.
This extended utility function~$ \hat u$ combines 
a copy of the previously constructed function $Y \owns y \mapsto u(y) \in \R $
with a supplementary function that assigns a utility value
 \( \hat \gamma (x) := \hat u( T_{> x} ) = \hat u( T_{ \ge x} ) \in \R \)
to each cut node $x \in X$ 
and so to each associated continuation subtree $ T_{>x} $
in the bounded domain $ \mathcal T^S (Y \cup \R )$ of trees that can be modelled. 

After a real-valued subjective evaluation 
has been attached to each truncation node, 
the description of Bayesian rational behaviour for ordinary decision trees $T$
that was provided in Section \ref{ss:valueSubtrees} can be extended 
to the typical truncated tree $ \hat T$.
Furthermore, the results cited in Section \ref{ss:justBayes} 
that have been used to justify Bayesian rationality obviously 
extend to truncated trees in which each truncation node 
is given a real-valued subjective evaluation.

\section{Enlivening Decision Trees} \label{s:enlivNodes}
\subsection{An Enlivened Deliberative Process} \label{ss:delib}

Recall the description in Section \ref{s:bayesRat} 
of Bayesian rational behaviour in finite decision trees.
In Section \ref{s:truncTrees}, that description was extended 
to truncated decision trees in which subjective evaluations 
are attached to any truncation node.
Including truncated decision trees suggests 
boundedly rational multi-step decision procedures,
along the lines that were informally discussed in Section \ref{ss:decProced}.
Recall too that each such procedure cycles between 
constructing the relevant continuation decision tree 
just after a decision has been made,
before moving to a new decision model which may enrich that subtree,
and then finally making a decision 
at the initial decision node of that new tree.
That decision, of course, takes the agent to the beginning of a new cycle. 
Recall too the lesson of Section \ref{ss:introEnliv} concerning 
the potential benefits to the agent of considering enlivened decision trees
which allow the possible implications of anticipating future model revisions 
to be exploraed.  

Accordingly, it is time to introduce enlivenment formally 
into the decision trees, including truncated trees, 
that were discussed in Sections \ref{s:bayesRat} and \ref{s:truncTrees}.
Both enlivenment and truncation will be regarded 
as parts of a deliberative process that, between successive decision nodes, 
modifies the agent's bounded decision model whose general form is, 
as discussed in Section \ref{ss:menuConseq}, 
a truncated decision tree \( \hat T \in \mathcal T^S ( Y \cup \R ) \).
As will be discussed in Section \ref{ss:enlivDomain},
enlivenment may affect both the consequence domain $Y$
and the domain $S$ of uncertain states of the world.
Accordingly, from now on we let both $Y$ and $S$ denote the current domains 
which apply at the current decision node.

In fact, the deliberation process typically pauses 
immediately before any decision node at which the agent is committed 
to a decision which is about to carried out.
Then, unless there are no more moves to make 
in the agent's decision tree model, 
the pause is only temporary; the enlivened deliberative process 
which describes the agent's decision procedure 
in any truncated decision tree $ \hat T$ 
will resume after the pause until the agent is once again 
about to make a move at a later decision node $n' \in N$.
The later node $n'$ is the initial node of a revised bounded decision model 
in the form of a continuation subtree $ \hat T_{ \ge n'} $.

An obvious key part of any deliberative process, whether enlivened or not, 
involves pruning off parts of the decision tree model 
which the agent knows can never be reached
because they have been excluded either by the agent's own earlier decisions, 
or by what the agent has learned 
about how one or more lotteries have been resolved.
This pruning process, of course, 
will typically release some of the agent's modelling resources.
As discussed in Section \ref{ss:decProced}, 
these resources can then be re-focused
on enlivening what remains of the previous bounded model,
which will typically benefit the agent,

\subsection{Enlivening Incomplete Decision Models} \label{ss:incompleteModels}

We start by distinguishing modelled from non-modelled enlivenment.
Consider again the Homeric myth of Odysseus and the Sirens 
that was discussed in Section \ref{s:homer}. 
The na\"\i ve sailors whose bones littered the meadow on the Sirens' island,
once they got within earshot of the Sirens' singing,
had faced a surprising and so unmodelled enlivening event.
By then it was already too late to escape their grisly fate; 
the sailors lacked the willpower to sail away from the island.
With Kirke's excellent and detailed advice, however, 
Odysseus was able to develop a much more accurate enlivened model 
of the Sirens' powers, and of what he could do to hear the Sirens 
and yet survive with his crew to continue their voyage.

The enlivenment phenomenon that we discuss here falls between 
the two cases of a modelled and an unmodelled enlivenment.
On the one hand, there is nothing that can be said about enlivenment 
which is not modelled at all; 
like that faced by the na\"\i ve sailors who perished on the Sirens' island.
Indeed, without at least a minimal model of enlivenment, 
the agent lacks any logical framework to allow for its possibility.
On the other hand, if the possibility of enlivenment 
is fully and properly modelled, that just changes the relevant decision tree 
and removes any need to discuss enlivenment.

So we focus on an intermediate case when the agent recognizes
the possibility of enlivenment and even models it, but only incompletely.
The resulting incomplete model is what emerges 
from whatever modelling procedure is used in the all too typical case 
when an ideal decision tree that treats fully
any potential enlivenment becomes impractically complex.
So what would otherwise be the ideal decision tree has to be truncated, 
and in a way that cannot, of course, be modelled satisfactorily.
That leaves open the possibility that interests us, 
which occurs when the decision maker is willing to recognize 
the possibility of revising the decision model, 
and so contemplates how to enliven the original decision tree.

Here, therefore, we consider how the agent's decision model 
will typically undergo an enlivening process 
in the form of a sequence of adjustments over time, 
just as the successive models revealed to Odysseus did 
while he was being advised by Kirke.
In that example, however, all the enlivenment took place even before 
Odysseus and his crew set sail again after Odysseus's encounter with Kirke.
Indeed, given a general decision tree, the agent may revise the current model 
due to dissatisfaction with the consequences of the decision it recommends.
In this connection, it may be worth repeating 
what we wrote in Section \ref{ss:boundedModels} emphasizing that,
unlike Simon's concept of satisficing which involves looking 
for a satisfactory decision within a given decision model,
the approach advocated here involves looking for a superior but bounded decision model 
which includes a feasible decision offering better consequences.  

More generally, however, 
the enlivening process will be interrupted and even disrupted
as the agent makes an actual decision that has been modelled in the decision tree.
Then any subsequent revisions should be to the agent's new refined model.
This is what remains of the original continuation decision tree 
given any subsequent moves the agent has made, 
as well as any lottery outcomes that have been observed.
Or, in the case when the agent is playing a game like Chess, 
the enlivening process may be interrupted and disrupted 
by observing an opponent's move. 

\subsection{Two Kinds of Minimal Enlivenment} \label{ss:2kinds}

Consider an agent whose decision procedure starts 
by facing a given finite truncated decision tree $ \hat T$ 
which we call the \emph{base tree}, with set of nodes~$N$ and set of edges $E$.
We assume that $ \hat T$ belongs to the domain \( \mathcal T^S (Y \cup \R ) \)
of decision trees specified in Definition \ref{def:treeDomain} 
of Section \ref{ss:decTree}, where~$Y$ is the consequence domain 
and $S$ is the domain of uncertain states of the world.
The process of enlivenment we consider starts with this base tree.
It then puts together a composition of successive enlivenments, 
all of which are \emph{minimal} 
in the sense that as much as possible of the tree $ \hat T$ is left unaffected.

Note first that if any enlivenment were to occur 
before the initial node $ n_0 $ of the tree $ \hat T$, 
then the agent has time to revise any previous decision 
so that it is appropriate in the enlivened tree.
So the tree $ \hat T$ could and should already have been redefined 
to include that enlivenment.
That redefinition, of course, leaves us with an unenlivened tree.

The first kind of minimal enlivenment, 
which we may call \emph{terminal node enlivenment},
takes place at a single terminal node.
Enlivenment replaces this terminal node, 
whether or not it is a truncation node,
by a decision, chance, or event node which is the initial node 
of a new finite continuation subtree.

\begin{figure} [hbt]
\begin{center}

\begin{picture}(260, 30)(25, 0)
	\put(20, -2){Subtree $ \hat T_{ \ge n}$:}
	\put(112, 0){ \circle* {8}}
	\put(110, 11){$n$}
	\put(115, 0){ \vector (1, 0){100}}
	\put(150, 8){$n \to m$} 
	\put(220, 0){ \circle* {8}}
	\put(218, 11){$ m$}
	\put(221, 0){ \vector (1, 0){20}}
	\put(250, -2){subtree $ \hat T_{ \ge m} $}
\end{picture}
\vspace{-1ex}

\end{center}
\caption{The Directed Edge $n \to m$ of the Truncated Tree $ \hat T = (N, E)$} 
	\label{fig:oneEdge}
\end{figure}
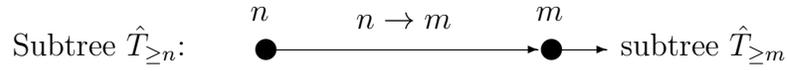

The second kind of minimal enlivenment, 
which we may call \emph{edge enlivenment}, is rather more complicated.
It is based on a single directed edge $n \to m$ 
of the truncated tree $ \hat T = (N, E) $, 
as illustrated in Figure \ref{fig:oneEdge}; 
all other edges of~$ \hat T$, including the nodes $n$ and $m$ 
at the beginning and end of $n \to m$, remain unchanged.
The initial node $n$ of the edge $n \to m$, of course, is the initial node 
of the continuation subtree $ \hat T_{ \ge n}$ whose initial node is $n$.
Upon reaching the terminal node $m$ of the edge $n \to m$, 
the agent before enlivenment faces 
the continuation decision tree $ \hat T_{ \ge m}$ whose initial node is~$m$.

\vspace{-3ex}

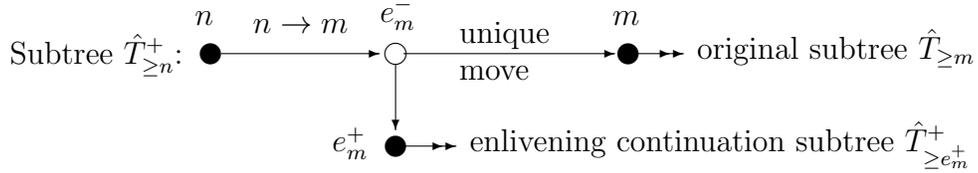
\begin{figure} [hbt]
\begin{center}
\vspace{-18ex}
\begin{picture}(300, 120)(0, -20)

	\put(-30, -4){Subtree $ \hat T^+_{ \ge n} $:}
	\put(42, 0) { \circle* {8}}
	\put(40, 11){$n$}
	\put(46, 0){ \vector (1, 0){60}}
	\put(62, 8){$n \to m$}
	\put(112, 0){ \circle {8}}
	\put(110, 12){$ e^-_m $}
	\put(115, 0){ \vector (1, 0){80}}
	\put(140, 4){unique}
	\put(140, -10){move}
	\put(200, 0){ \circle* {8}}
	\put(198, 11){$m$}
	\put(201, 0){ \vector (1, 0){15}}
	\put(216, 0){ \vector (1, 0){5}}
	\put(230, -2){original subtree $ \hat T_{ \ge m}$}
	\put(112, -4){ \vector (0, -1){25}}
	\put(92, -37){$ e^+_m $}
	\put(112, -35){ \circle* {8}}
	\put(115, -35){ \vector (1, 0){15}}
	\put(130, -35){ \vector (1, 0){5}}
	\put(144, -37){enlivening continuation subtree $ \hat T^+ _{ \ge e^+_m }$}

\end{picture}

\end{center}
\vspace{1ex}
\caption{A Minimal Enlivenment of the Directed Edge $n \to m$}
\label{fig:enlivNode}
\end{figure}

Next, Figure~\ref{fig:enlivNode} shows the effect 
on the specific directed edge $n \to m$ of the minimal enlivenment 
consisting of:
\begin{enumerate}
	\item first, inserting into $n \to m$  
	the additional \emph{enlivenment node} $ e^-_m $ that is an event node,
	where the outcome of a horse lottery, 
	as defined in Section \ref{ss:horseLotts}, 
	determines which node from the pair $\{ m, e^+_m \}$ succeeds~$ e^-_m $;
	\item second, adding the extra enlivenment edge $ e^-_m \to e^+_m $ 
	 ending at the extra \emph{post-enlivenment node} $ e^+_m $ which is:
		\begin{itemize}
			\item apart from~$m$ itself, 
			the only other immediate successor of $ e^-_m $;
			\item the initial node 
			of the \emph{enlivening continuation subtree} 
			$ \hat T^+ _{ \ge e^+_m }$.
		\end{itemize}
\end{enumerate}

Thus, the horse lottery that occurs at the event node $ e^-_m $ 
determines whether an uncertain deviation 
from the path $n \to e^-_m \to m$ in the original truncated tree $ \hat T$ 
to the alternative path $n \to e^-_m \to e^+_m $ does or does not occur.
Specifically, as illustrated in Figure \ref{fig:enlivNode}:
\begin{enumerate}
	\item First, if no deviation occurs, 
	then the relevant immediate successor of $ e^-_m $ 
	in the minimal enlivenment $ T^+_{ \ge n} $ of $ \hat T_{ \ge n} $ 
	is the initial node $m$ 
	of the original unenlivened continuation subtree $ \hat T_{ \ge m}$ 
    that emanates from $m$ in the original unenlivened tree $ \hat T$.
    \item Alternatively, if a deviation does occur, 
    then the relevant immediate successor of $ e^-_m $ 
    is the \emph{post-enlivenment node} $ e^+_m $.
	This is the initial node of a different 
	finite continuation subtree $ \hat T^+_{ \ge e^+_m }$ that gets appended 
	to $ \hat T_{ \ge n} $ at $ e^+_m $ in the process 
	of enlivening $ \hat T_{ \ge n} $ to the new tree $ \hat T^+_{ \ge n} $.
\end{enumerate}

\subsection{Three Levels of Enlivenment} \label{ss:3levels}

The minimal enlivenments described in Section \ref{ss:2kinds} 
are only the lowest first level.
A second level results from \emph{basic enlivenments} 
where each terminal node of $ \hat T$ 
may be given one terminal node enlivenment, 
whereas each edge of~$ \hat T$ may be given one edge enlivenment.
By definition of decision tree, both the original decision tree 
and each new continuation decision tree that emerges from a basic enlivenment
have a finite collection of terminal nodes and edges.
So any basic enlivenment always results in a finite enlivened tree.

In the account of the Homeric example in Section \ref{s:homer}, however,
most of the enlivenments were not basic.
Instead the decision tree that Kirke described to Odysseus 
was enlivened in several successive stages, 
of which only the first stage involves a basic enlivenment of the base tree.
Indeed, after this first stage, all subsequent enlivenments 
were of parts of the enlivened decision tree
that had only been added as part of a previous enlivenment.
This makes evident the possibility 
of a third level of \emph{recursive enlivenment},
which occurs when parts of the tree that have been added to the base tree 
during the enlivening process can themselves become enlivened at a later stage.
As another example, anybody who has ever played Chess 
at any level beyond the most basic 
also knows that recursive enlivenment affects a player's evolving understanding 
of the game being played, and so of how to evaluate any position in that game.

Note that even a minimal enlivenment 
adds at least one node to the original tree.
And that, by definition, any basic enlivenment 
combines a finite collection of minimal enlivenments. 
Recursive enlivenment, however, can proceed indefinitely 
and produce a tree with infinitely many nodes.
In this paper, however, we consider only finite decision trees. 
This limits our analysis to enlivenment processes 
that end after finitely many stages.

\section{Bayesian Rationality in Enlivened Trees} \label{s:ratEnliv}
\subsection{Enlivened Domains of Consequences and States} \label{ss:enlivDomain}

In Sections \ref{s:bayesRat} and \ref{s:truncTrees} 
we considered a fixed consequence domain $Y$ 
together with a fixed domain $S$ of uncertain states of the world.
Then we discussed Bayesian rationality and subjective evaluations
for the corresponding fixed domain $ \mathcal T^S (Y \cup \R )$ 
of finite decision trees $ \hat T$, including their continuations and truncations.

Homer's example of Odysseus and the Sirens, however, 
as discussed in Section \ref{s:homer}, should alert us to the possibility 
that either or both of $Y$ and~$S$ may be expanded by enlivenment.
After all, consider what happens as a result of the three stages of enlivenment
as one passes from the first na\"\i ve sailor's tree in Figure \ref{f:naive}
to Kirke's final model which appears 
as the fourth tree in Figure \ref{f:odyss2}.
The obvious domain of uncertain states expands 
to include the possibility that Sirens exist.
And the obvious consequence domain expands to include the grisly possibility 
that Odysseus and all his crew die on the Sirens' island,
as well as the alternative and much superior possibility 
that Odysseus hears the Sirens, yet he and his crew all manage to get away.

Accordingly, we allow any step of an enlivenment process to expand 
either or both of the consequence domain $Y$ and state domain $S$.
Yet, as we just argued in Section \ref{ss:3levels}, 
in restricting attention to finite decision trees, 
we necessarily limit our analysis to enlivenment processes 
that end after finitely many steps.
It follows that any permitted process must terminate 
with both a limiting consequence domain $ Y^+ $ 
and a limiting state domain~$ S^+ $.

\subsection{Rationality with Enlivened Utility Functions} \label{ss:enlivUtils}

Recall that two distinct fixed consequences $ \underline y, \bar y \in Y$ 
were used in Section \ref{ss:constructBUF} in order to construct 
the unique normalized Bernoulli utility function 
that satisfies \eqref{eq:normBUF}.
Notice that the new domain $ \mathcal T^{ S^+ } ( Y^+ \cup \R )$ 
of enlivened decision trees and their truncations must include, 
for each consequence $y \in Y^+ $, 
the set $ \mathcal T( \Delta ( \{ \underline y, \bar y, y \} )) $ 
of risky finite decision trees, 
without any event nodes, horse lotteries, or uncertain states, 
whose roulette consequence lotteries are restricted to probability mixtures
of the three degenerate lotteries
 $ \delta _{ \underline y} $, $ \delta _{ \bar y} $, and $ \delta _y $.
Then the construction set out in Section \ref{ss:constructBUF} can be repeated 
to determine the utility $ u( y^+ )$ 
of each extra consequence $ y^+ \in Y^+ \setminus Y$.
The result is a normalized enlivened Bernoulli utility function 
\begin{equation} \label{eq:enlivenBUF}
	Y^+ \owns y^+ \mapsto u^+ ( y^+ ) \in \R
\end{equation}
that satisfies \eqref{eq:normBUF} 
not only for consequences $y$ in the original domain $Y$, 
but also for any extra consequences $ y^+ $ 
in the entire exlivened domain $ Y^+ $.
Moreover, the expectation of $ u^+ $ defined by \eqref{eq:enlivenBUF}
will represent the agent's extended preference relation $ \succsim ^+ $ 
on the whole lottery domain $ \Delta ( Y^+ )$.
Also, because the two particular consequences $ \underline y, \overline y$ 
used in the earlier normalization \eqref{eq:normBUF} are in $ Y^+ $ 
as well as in $Y$, we can the impose the obvious counterpart 
\begin{equation} \label{eq:normEnlivenBUF}
	u^+ ( \underline y) 
	= \underline u \quad \text{and} \quad u^+ ( \overline y) = \overline u
\end{equation}
of that earlier normalization.
The resulting function \eqref{eq:enlivenBUF} 
will then extend $Y \owns y \mapsto u(y) \to \R $ 
to the expanded domain $ Y^+ $.
Similarly, by considering the domain $ \mathcal T^{ S^+ }( Y^+ \cup \R )$ 
instead of $ \mathcal T^S ( Y^+ \cup \R )$, 
the construction of the subjective probabilities $ \mathbb P(s) $ 
of each state $s \in S$ can be extended to a construction 
of the enlivened subjective probabilities $ \mathbb P^+ (s) $ 
of each state $s \in S^+ $.

With these constructions 
of the enlivened consequence domain $ Y^+ $ and state domain $ S^+ $,
as well as of the normalized enlivened Bernoulli utility function on $ Y^+ $
satisfying \eqref{eq:enlivenBUF} and \eqref{eq:normEnlivenBUF},
and of the enlivened subjective probabilities $ \mathbb P^+ (s) $ 
of each state $s \in S^+ $, the justification 
offered in Section \ref{ss:justBayes} for Bayesian rational behaviour 
in decision trees $T$ that belong to the domain $ \mathcal T^S (Y \cup \R )$
obviously extends to Bayesian rational behaviour in enlivened decision trees $ T^+ $ 
that belong to the domain $ \mathcal T^{ S^+ }( Y^+ \cup \R )$.

\subsection{Limitations of Bayesian Rationality} \label{ss:limitsBayesian}

A concept of rationality that is more philosophically refined 
than mere Bayes\-ian rationality would presumably require a rational agent 
to use some kind of ``normatively justified'' Bernoulli utility function
defined on the consequence domain,
together with some kind of ``normatively justified'' subjective probabilities 
over uncertain states of the world.%
\footnote{As a referee has kindly suggested, this paper 
is not the place to embark on a philosophical discussion of what precisely 
the phrase ``normatively justified'' could and should mean.
A reader who wishes to get some idea of the depth 
to which such a discussion should go may want to consult authoritative works 
such as Gibbard (1990, 2003, 2012), Broome (2013, 2021), or Bratman (2018).}
Thus, this richer concept of rationality 
would go beyond mere Bayesian rationality.

Bayesian rationality in enlivened trees is no less limited.
Indeed, full rationality should require the agent's subjective evaluations 
at terminal evaluation nodes to be normatively justified, 
in addition to the agent's normalized Bernoulli utility function 
and subjective probabilities.
Of course, the agent's estimates 
of the relevant subjective probabilities and subjective evaluations 
may well be improved by procedures 
such as Monte Carlo tree search or scenario planning,
as considered in Section \ref{ss:stockfishChess}.

\section{More Examples and Extensions} \label{s:apps}
\subsection{The Precautionary Principle and Option Values} \label{ss:precPrinc}

The European Commission has provided 
an official definition of the precautionary principle:%
\footnote{\url{https://
	eur-lex.europa.eu/EN/legal-content/glossary/precautionary-principle.html}}

\begin{quote}
    The precautionary principle is an approach to risk management, where, 
    if it is possible that a given policy or action 
    might cause harm to the public or the environment 
    and if there is still no scientific agreement on the issue, 
    the policy or action in question should not be carried out. 
    However, the policy or action may be reviewed 
    when more scientific information becomes available.
    
    The concept of the precautionary principle was first set out 
    in a European Commission communication adopted in February 2000, 
    which defined the concept and envisaged how it would be applied.
    
    The precautionary principle may only be invoked 
    if there is a potential risk and may not be used to justify arbitrary decisions.
\end{quote}

But the concluding sentences of Sunstein (2003) draw attention 
to some serious possible limitations of the principle:%
\footnote{See Sunstein (2003), pp.\ 1057--1058.
	For related criticisms see Sunstein (2005). 
	Many of these criticisms are addressed, at least partially, in Steele (2006).}

\begin{quote}
    My goal here has been \ldots to explain 
    the otherwise puzzling appeal of the precautionary principle 
    and to isolate the strategies that help make it operational. 
    At the individual level, these strategies are hardly senseless, 
    especially for people who lack much information or who do the best they can 
    by focusing on only one aspect of the situation at hand. 
    But for governments, the precautionary principle is not sensible, 
    for the simple reason that once the viewscreen is widened, 
    it becomes clear that the principle provides no guidance at all. 
    A rational system of risk regulation certainly takes precautions. 
    But it does not adopt the precautionary principle.
\end{quote}

As discussed in Gollier and Treich (2003), for example,
whatever decision-theoretic principles underlie the precautionary principle
seem to have emerged from work such as that by Arrow and Fisher (1974) 
on environmental preservation and irreversible investments. 
On page 317, while considering a two-period model, 
Arrow and Fisher write that

\begin{quote}
 	Given an ability to learn from experience, 
	underinvestment can be remedied before the second period, 
	whereas mistaken overinvestment cannot, 
	the consequences persisting in effect for all time.
\end{quote}

The concluding sentence of their paper (p.\ 319) reads as follows:%
\footnote{Arrow and Fisher add a footnote citing Hart (1942),
whose work Arrow had earlier often credited with arousing his interest 
in the economics of risk and uncertainty.}

\begin{quote}
	Essentially, the point is that the expected benefits of an irreversible decision 
	should be adjusted to reflect the loss of options it entails.
\end{quote}

Arrow and Fisher's use of the phrase ``mistaken overinvestment'' 
makes it clear that they are considering a case where investment decisions
can be at best boundedly rational.
This could be because a typical decision procedure 
may well overlook some of the possible consequences of a decision, 
particularly concerning the environment.
Some of these consequences, if not completely irreversible, 
could at least be very expensive to reverse.
Then one could argue for using a decision procedure 
that makes some allowance for the impossibility of ever modelling in full 
all the fundamental consequences that may be relevant.
One way to make such an allowance is to enliven the decision tree being analysed.
Then the tree can allow as far as practicable for the possibility 
that at least some important consequences which it omits
may have to be included in a later revised analysis.
The precautionary principle seems to be one of many possible ways to do this.

\subsection{Transformative Experiences} \label{ss:transExpers}

The concept of transformative experiences arose in philosophy 
thanks to Paul (2014, 2015a, b, c), 
though similar ideas were discussed earlier in Ullmann-Margalit's (2006) paper 
on ``big decisions'' whose first characteristic (p.\ 158) 
is that they must be ``transformative, or `core affecting'{\,}''.
As Paul (2015b, p.\ 761) writes:
\begin{quote}
    a transformative experience is \ldots both radically new to the agent 
    and changes her in a deep and fundamental way;
    there are experiences such as becoming a parent, discovering a new faith,
    emigrating to a new country, or fighting in a war. \ldots
    
    An \emph{epistemically} transformative experience is an experience 
    that teaches you something you could not have learned 
	without having that kind of experience. 
    Having that experience gives you new abilities to imagine, recognize, 
    and cognitively model possible future experiences of that kind.
    A \emph{personally} transformative experience changes you 
    in some deep and personally fundamental way, for example, 
    by changing your core personal preferences 
    or by changing the way you understand your desires 
    and the kind of person you take yourself to be. 
    A \emph{transformative experience}, then, is an experience 
    that is both epistemically and personally transformative.
    
    The main problem with transformative decisions 
    is that our standard decision models break down when we lack 
    epistemic access to the subjective values for our possible outcomes.
\end{quote}

As two philosophically prominent examples of transformative experiences,
she considers the decisions of whether to become a vampire (!) 
or to have a child.
See also the discussion 
by Pettigrew (2015, 2020), Barnes (2015), and Campbell (2015), 
as well as later by writers 
such as Bykvist and Stef\'ansson (2017) and Randell (2023),
as well as Sunstein and Paul (2024).

Consider any decision that determines 
whether or not the agent undergoes a transformative experience.
The claim being made in this paper is that the decision can be modelled 
in an enlivened decision tree whose potential consequences 
include the results of the experience.
In the case of an experience 
which is classified as epistemically but not personally transformative,
the transformation can be modelled by allowing an enlivenment that affects 
only the domain of relevant uncertain states of the world 
and/or the probabilities of those states.

An experience which is classified as personally transformative, however,
may transform the agent's personal identity in ways that most philosophers 
have typically regarded as posing serious conceptual challenges 
to normative decision theory.
Pettigrew (2020) is one philosopher 
who has made a notable attempt to address these challenges.
His work deserves fuller discussion elsewhere.
Here I merely note that some economists may want to regard 
personally transformative experiences as leading to nothing more
than changes in a very broadly conceived personal characteristic 
which can vary within a specified domain of such characteristics
that encompasses, for example, all possible changes in personal identity.
See, for example, ideas discussed 
by Fleurbaey and Hammond (2004, especially Section 7), who attempt to go 
beyond Kolm's (1972, 1994, 2002) work on ``fundamental preferences''
while meeting critiques such as Broome's (1993, 1994).%
\footnote{A very recent mathematically sophisticated contribution 
to the relevant literature is Pivato (2025).}
One possibility that may be worth exploring is that an enlivened decision tree
might be used to explore the implications of unforeseen changes 
in the relevant domain of personal characteristics, 
including the relevant consequences of weighty ethical decisions.

\subsection{Decision Trees with Timed Consequence Nodes} \label{ss:menuConseqs}

In Hammond and Troccoli Moretti (2026) we consider finite decision trees which,
in addition to decision, chance, event, and terminal nodes, 
also have timed consequence nodes.
Each timed consequence node is assumed to have a unique immediate successor.

Given the consequence domain $Y$, for each $m \in \N$ let $ Y^m $ denote 
the Cartesian product of $m$ copies of $Y$.
Then let $ \mathbf Y(Y) $ denote the domain
of all possible timed consequence streams which, for some $m \in \N$,
take the form $ \mathbf y = ( t_j, y_j ) _{j = 1}^m \in \R^m \times Y^m $,
where the sequence $( t_j ) _{j = 1}^m $ of times 
satisfies $ t_1 < t_2 < \ldots < t_m $.
With this construction, any deterministic truncated decision tree $ \hat T$ 
with timed consequence nodes can be reduced 
to an \emph{equivalent} decision tree $ T^* \in \mathcal T( \mathbf Y(Y)) $ 
in which:
\begin{enumerate}
    \item at each terminal node $n \in N^t $, which is a consequence node,
    the assigned terminal consequence $ \gamma (n) \in Y$ is replaced 
    by the unique timed consequence stream $ \mathbf y(n) \in \mathbf Y(Y) $ 
    that ends at node $n$; 
    \item each timed consequence node is then eliminated, 
    and the two directed edges that immediately precede and succeed 
    the eliminated node are amalgamated into a single directed edge.
\end{enumerate}
In addition to deterministic decision trees, 
a similar reduction is possible for all decision trees 
that have chance and/or event nodes.

In a decision tree that includes timed consequence nodes, 
the eventual consequence of reaching a terminal node $n$ is,
not a lottery over single consequences at node $n$, 
but rather a lottery over the stream of consequences and consequence lotteries 
which accumulate along the unique path through the decision tree 
that ends at node~$n$.
The implications of allowing timed consequence nodes are then routine 
unless the consequence of reaching any consequence node $n$ 
includes a ``menu consequence'' which depends 
on what consequence lotteries are feasible 
in the continuation decision tree $ \hat T_{ \ge n}$ emanating from node~$n$.
Even then, however, the earlier results summarized in Section \ref{ss:justBayes}
that justify Bayesian rationality remain valid; all that changes,
in order to accommodate each consequence stream that includes a menu consequence,
is that the domain of relevant consequence streams becomes much richer.

These changes can be significant, however.
For example, they allow scope for preferences for flexibility,
as in Koopmans (1950, 1964) and Kreps (1979).
They also allow preferences for the timing of resolution of uncertainty, 
as in Kreps and Porteus (1978). 
Other possibilities are menu consequences that reflect preferences over lotteries 
such as those considered by Machina (1989) and by Gilboa and Schmeidler (1989),
which cannot be represented by an expected utility function
over the domain of timed consequence streams.
In addition, menu consequences allow recursive preferences 
for intertemporal consumption streams 
of the kind that Epstein and Zin (1989) first considered. 

Indeed, for trees with timed consequence nodes 
as well as decision, chance and event nodes, 
there is an obvious extension of the rules set out in the four equations 
\eqref{eq:termVal}, \eqref{eq:chanceVal}, \eqref{eq:eventVal}, 
and \eqref{eq:maxValue} of Section \ref{ss:valueSubtrees}.
This extension treats the case when node $n$ is a consequence node,
in which case the timed consequence attached to node $n$ is prepended 
at the beginning of each consequence stream $ \mathbf y$ 
which arises in the continuation subtree $ \hat T_{ \ge n}$.
With this extension, the description in Section \ref{s:bayesRat}
of Bayesian rationality for enlivened decision trees is still valid.

A minor complication does arise, however, in describing Bayesian rationality 
with subjective evaluations in enlivened but truncated decision trees.
The issue is how to truncate a non-trivial stream of consequences 
and consequence lotteries that accumulate along a path 
through the decision tree.
The obvious remedy, following the ideas in Section \ref{s:truncTrees}, is to attach 
to each cut or truncation node $x$ a real-valued combined subjective evaluation 
of the consequence stream that has accumulated before the truncation,
along with an allowance for the unmodelled continuation decision tree 
which has been cut off.
Together, the effect should be an estimate of what subjective evaluation 
would have emerged if the backward recursion described 
in Section \ref{ss:truncatedEvals} could have been be carried out in full.
In economic applications, this approach could accommodate in principle
the kind of decision problem which consumers and producers both face 
in dynamic models such as those considered by, 
inter alia, Hicks (1946) and Myerson (1983).

\subsection{Reverse Bayesianism} \label{ss:revBayes}

``Reverse Bayesianism'' was described in the series of joint papers 
by Karni and Vier\o\ (2013, 2015, 2017).
For the general finite decision trees considered here, 
reverse Bayesianism is the result saying that, 
in any enlivened decision tree $ T^+ $,
if you condition the probabilities of different consequences 
on the event that enlivenment does not occur, 
then the result should be the corresponding probabilities 
in the original unenlivened decision tree $ \hat T$.
More specifically, the original subjective probabilities $ \mathbb P(s) $ 
can be recovered from the enlivened subjective probabilities $ \mathbb P^+ (s) $
by considering the conditional probabilities 
given the event that $s$ belongs to the unenlivened state space $S$,
rather than to the enlivened state space $ S^+ $.

Some recent criticisms of reverse Bayesianism by Steele and Stef\'ansson (2021a, b) 
are based on Bradley's (2017, p.\ 257) description of the principle:
\begin{quote}
 	\ldots in cases where we become aware of prospects that are inconsistent 
	with those that we previously took into consideration
	\ldots [reverse Bayesianism requires] that we should extend 
	our relational attitudes to the new set 
	in such a way as to conserve all prior relational beliefs.
\end{quote}
Now, a crucial feature of all decision trees considered in this paper, 
including those that are enlivened but truncated,
is that they follow Hammond (1988a, b; 1998a, b; 1999, 2022) 
in having only positive probabilities attached to all chance or event nodes.  
Furthermore, truncations take the special form 
described in Section \ref{s:truncTrees}, 
where continuation subtrees are removed 
but their initial nodes remain as terminal nodes of the truncated tree.
It follows that the revised beliefs which Steele and Stef\'ansson present 
as exceptions to reverse Bayesianism point to the theoretical limitations
of considering only belief revisions that result 
from enlivenment and permitted truncation of decision trees 
where all probabilities are positive.

\section{Enlivened Bounded Rationality} \label{s:concs}

The widely quoted aphorism due to the statistician George Box 
that was reproduced at the head of Section \ref{ss:boundedModels}
should remind us of the inevitable limitations which will arise 
in any formal model of all but the most trivial decision problems.
Indeed, examples that extend in time from Homer's \textit{Odyssey} 
to modern algorithms for playing Chess demonstrate that, 
for an agent who has one or more decisions to make, 
the usefulness of any model is all too likely to be temporary.
This paper has begun an investigation of how to adapt 
the concept of Bayesian rationality so it can be used by an agent who recognizes 
and makes a serious but inevitably incomplete and imperfect attempt 
to deal with the fundamental difficulty that enlivenment creates.

Specifically, it is argued that the best decision procedure 
which such an agent can follow is:
(i) immediately before reaching any decision node, 
and then allowing for possible enlivenment as far as possible, 
construct a subjective truncated probabilistic model 
based on an estimated expected value of what the ultimate ex post evaluation 
of each possible decision should be; 
(ii) at any decision node, make a decision 
that maximizes this estimated expected value;
(iii) before each successive decision, apply appropriate enlivenments 
of the kind considered in Section \ref{s:enlivNodes} 
to the relevant continuation of the model used in stage (i) 
in order to recognize any of its deficiencies 
which may have become apparent since the latest previous decision.

\newpage

\begin{center}
	\textbf{Acknowledgements}
\end{center}

The research reported here was initially supported 
from April 2007 to March 2010 
by a Marie Curie Excellence Chair funded by the European Commission 
under contract number MEXC-CT-2006-041121.
Many thanks also to Kenneth Arrow, Ken Binmore, and Burkhard Schipper 
for their helpful suggestions during these early years,
to Stan Metcalfe for increasing my appreciation of Shackle's original ideas,
as well as to Marcus Miller and Joanne Yoong for enlivened discussions,
while absolving them of all responsibility for my errors or omissions.

Many thanks also for their patient attention and encouragement to audiences at:
the GSB/Economics Department theory seminar 
at Stanford University (March 2007);
the London School of Economics conference on preference change (May 2009);
the University of Warwick (November 2010);
CORE at the Universit\'e Catholique de Louvain (September 2011);
the Hausdorff Center for Mathematics in Bonn (August 2013);
the Universit\`a Cattolica del Sacro Cuore in Milan (October 2013);
the conference on unawareness in game theory 
at the University of Queensland (February 2014);
the conference on game theory and applications 
at the University of Bristol (May 2016);
the Microeconomics Work in Progress online seminar 
at the University of Warwick (March 2021); 
the workshop on ``Issues in Dynamic Decision Theory'' 
at the University of Konstanz (July 2023);
the Centre d'\'Economie de la Sorbonne in Paris (November 2023);
the IX Hurwicz Workshop on Mechanism Design Theory 
at the SGH Warsaw School of Economics and at the Banach Center
of the Institute of Mathematics 
of the Polish Academy of Sciences (December 2024);
a seminar at the Economics Department 
of the University of Liverpool (April 2025);
and finally at the RUD (Risk, Utility, and Decision) 2025 conference 
in Manchester (June 2025) 
as well as at the SAET (Society of Advancement of Economic Theory) conference 
on the island of Ischia, Italy (July 2025).

Recently, the prolonged process of producing the current version of this paper 
was greatly helped by discussions with Agust\'\i n Troccoli Moretti, 
my coauthor on a related project,
as well as with my colleagues Kirk Surgener and Pablo Beker at Warwick.
Last but far from least, 
this greatly improved version of an earlier CRETA working paper 
owes much to the helpful comments of two unusually diligent anonymous referees.

\newpage

\begin{center} \textbf{References}
\end{center}

\begin{description}
\small

\item Abramson, B. (1991) 
	\textit{The Expected-Outcome Model of Two-Player Games} 
	(San Mateo: Morgan Kaufman).
	
\item Anscombe, F.J., and R.J. Aumann (1963)
	``A Definition of Subjective Probability''
	\textit{Annals of Mathematical Statistics} 34: 199--205.
	
\item Arrow, K.J., and A.C. Fisher (1974) 
	``Environmental Preservation, Uncertainty, and Irreversibility'' 
	\textit{Quarterly Journal of Economics} 88 (2): 312--319.
	
\item Barnes, E. (2015) 
	``What You Can Expect When You Don't Want to be Expecting'' 
	\textit{Philosophy and Phenomenological Research} 91 (3): 775--786.
	
\item Bradley, R. (2017) \textit{Decision Theory with a Human Face} 
	(Cambridge: Cambridge University Press).
	
\item Bratman, M.E. (2018) \textit{Planning, Time, and Self-Governance: 
	Essays in Practical Rationality} (Oxford: Oxford University Press).
	
\item Broome J. (1993) ``A Cause of Preference Is Not an Object of Preference''
	 \textit{Social Choice Welfare} 10: 57--68.

\item Broome, J. (1994) ``Reply to Kolm'' 
	\textit{Social Choice Welfare} 11: 199--201.

\item Broome, J. (2013) \textit{Rationality Through Reasoning} 
	(Oxford: Wiley-Blackwell).

\item Broome, J. (2021) 
	\textit{Normativity, Rationality and Reasoning: Selected Essays} 
	(Oxford: Oxford University Press).

\item Bykvist, K., and H.O. Stef\'ansson (2017) 
	``Epistemic Transformation and Rational Choice'' 
	\textit{Economics and Philosophy} 33(1): 125--138.
	
\item Campbell, J. (2015) ``L.A. Paul's \textit{Transformative Experience}''
	\textit{Philosophy and Phenomen\-ological Research} 91(3): 787--793.

\item Dekel, E., B.L. Lipman, and A. Rustichini (2001)
	``Representing Preferences with a Unique Subjective State Space,''
	\textit{Econometrica} 69: 891--934.

\item Dekel, E., B.L. Lipman, A. Rustichini, and T. Sarver (2007)
	``Representing Preferences with a Unique Subjective State Space: 
	Corrigendum'' \textit{Econometrica} 75: 591--600.
	
\item Derbyshire, J. (2017) ``Potential Surprise Theory 
	as a Theoretical Foundation for Scenario Planning''
	\textit{Technological Forecasting and Social Change} 124: 77--87.
 
\item Dr\`eze, J.H. (1962) ``L'utilit\'e sociale d'une vie humaine'' 
	\textit{Revue Fran\c caise de Re\-cherche Operationelle} 23: 93--118,

\item Dr\`eze, J.H., and A. Rustichini (2004) 
	``State-Dependent Utility and Decision Theory''
	in S. Barber\`a, P.J. Hammond and C. Seidl (eds.)
	\textit{Handbook of Utility Theory, Vol.\ 2: Extensions}
	(Dordrecht: Kluwer Academic) ch. 16, pp. 839--892.
	
\item Earl, P.E., and B. Littleboy (2014) \textit{G.L.S.\ Shackle} 
	(Basingstoke, UK: Palgrave Macmillan).
	
\item Elster, J. (1979)
	\textit{Ulysses and the Sirens: Studies in Rationality and Irrationality}
	(Cambridge: Cambridge University Press).

\item Epstein, L.G., and S.E. Zin (1989) ``Substitution, Risk Aversion, 
	and the Temporal Behavior of Consumption and Asset Returns: 
	A Theoretical Framework'' \textit{Econometrica} 57 (4): 937--969.
	
\item Fleurbaey, M. and P.J. Hammond (2004) 
	``Interpersonally Comparable Utility''
	in S. Barber\`a, P.J. Hammond and C. Seidl (eds.)
	\textit{Handbook of Utility Theory, Vol.\ 2: Extensions}
	(Dordrecht: Kluwer Academic) ch.\ 21, pp. 1181--1285.

\item Gibbard, A. (1990) 
	\textit{Wise Choices, Apt Feelings: A Theory of Normative Judgment} 
	(Harvard University Press).

\item Gibbard, A. (2003) \textit{Thinking How to Live} 
	(Harvard University Press). 

\item Gibbard, A. (2012) \textit{Meaning and Normativity} 
	(Oxford University Press).

\item Gilboa, I., and D. Schmeidler (1989) 
	``Maxmin Expected Utility with Non-Unique Prior''
	\textit{Journal of Mathematical Economics} 18: 141--153.

\item Gollier, C., and N. Treich (2003) 
	``Decision-Making Under Scientific Uncertainty: 
	The Economics of the Precautionary Principle''
	\textit{Journal of Risk and Uncertainty} 27: 77--103.

\item Grant, S. and J. Quiggin (2015a) 
	``A Preference Model for Choice Subject to Surprise''
	 \textit{Theory and Decision} 79 (1): 167--180.
	 
\item Grant, S., J. Kline, P. O'Callaghan, and J. Quiggin (2015b) 
	``Sub-models for Interactive Unawareness''
	\textit{Theory and Decision} 79 (4): 601--613.
	
\item Halpern, J.Y. and L.C. R\^ego (2014) 
	``Extensive Games with Possibly Unaware Players'' 
	\textit{Mathematical Social Sciences} 70: 42--58.

\item Hammond, P.J. (1976) ``Changing Tastes and Coherent Dynamic Choice''
	\textit{Review of Economic Studies} 43: 159--173.
	
\item Hammond, P.J. (1988a) ``Consequentialism and the Independence Axiom'' 
	in B.R. Munier (ed.) \textit{Risk, Decision and Rationality 
	(Proceedings of the 3rd International Conference 
	on the Foundations and Applications 
	of Utility, Risk and Decision Theories)} 
	(Dordrecht: D. Reidel), pp. 503--516.

\item Hammond, P.J. (1988b) 
	``Consequentialist Foundations for Expected Utility''
	\textit{Theory and Decision} 25: 25--78.

\item Hammond, P.J. (1998a) 
	``Objective Expected Utility: A Consequentialist Perspective''
	in S. Barber\`a, P.J. Hammond and C. Seidl (eds.)
	\textit{Handbook of Utility Theory, Vol. 1: Principles}
	(Dordrecht: Kluwer Academic) ch. 5, pp. 145--211.

\item Hammond, P.J. (1998b) ``Subjective Expected Utility''
	in S. Barber\`a, P.J. Hammond and C. Seidl (eds.)
	\textit{Handbook of Utility Theory, Vol. 1: Principles}
	(Dordrecht: Kluwer Academic) ch. 6, pp. 213--271.

\item Hammond, P.J. (1999) ``Subjectively Expected State-Independent Utility 
	on State-Dep\-endent Consequence Domains'' 
	in M.J. Machina and B. Munier (eds.) 
	\textit{Beliefs, Interactions, and Preferences in Decision Making}
	(Dordrecht: Klu\-wer Academic), pp.\ 7--21.

\item Hammond, P.J. (2007) ``Schumpeterian Innovation 
	in Modelling Decisions, Games, and Economic Behaviour'' 
	\textit{History of Economic Ideas} XV: 179--195. 

\item Hammond, P.J. (2022) 
	``Prerationality as Avoiding Predictably Regrettable Consequences''
	\textit{Revue \'Economique} 73: 943--976.

\item Hammond, P.J. and A. Troccoli Moretti (2026) ``Prerationality 
	in Risky Decision Trees with Timed and Menu Consequence Nodes''
	University of Warwick, CRETA working paper (in preparation).

\item Hansen, L.P. and T. Sargent (2007) \textit{Robustness}
	(Princeton: Princeton University Press).
	
\item Hart, A. G. (1942) 
	``Risk, Uncertainty, and the Unprofitability of Compounding Probabilities''
	in O. Lange, F. McIntyre, and T. O. Yntema (eds.) 
	\textit{Studies in Mathematical Economics and Econometrics}
	(Chicago: University of Chicago Press).

\item Hicks, J.R. (1946) \textit{Value and Capital: 
	An Inquiry into Some Fundamental Principles of Economic Theory}
	(Oxford: Clarendon Press).

\item Jefferson, M. (2012) ``Shell Scenarios: What Really Happened 
	in the 1970s and What may be Learned for Current World Prospects''
	 \textit{Technological Forecasting and Social Change} 79 (1): 186--197.
	
\item Jefferson, M. (2014) 
	``The Passage of Time: Shackle, Shell and Scenarios''
	in Earl, P., and B. Littleboy (eds.) \textit{G.L.S.\ Shackle}
	(Basingstoke: Palgrave Macmillan) pp.\ 198--214.

\item Karni, E. (1985) \textit{Decision Making Under Uncertainty: 
	The Case of State-Depend\-ent Preference}
	(Cambridge, MA: Harvard University Press).

\item Karni, E. and M.-L. Vier\o\ (2013) 
	``{}`Reverse Bayesianism': A Choice-Based Theory of Growing Awareness''
	 \textit{American Economic Review} 103: 2790--2810.
	 
\item Karni, E. and M.-L. Vier\o\ (2015) 
	``Probabilistic Sophistication and Reverse Bayes\-ian\-ism''
	 \textit{Journal of Risk and Uncertainty} 50: 189--208.

\item Karni, E. and M.-L. Vier\o\ (2017) ``Awareness of Unawareness:
	A Theory of Decision Making in the Face of Ignorance''
	\textit{Journal of Economic Theory} 168: 301--328.

\item Kolm, S.-C. (1972) \textit{Justice et \'equit\'e} 
	\textit{Centre National de la Recherche Scientifique}.

\item Kolm, S.-C. (1994) ``The Meaning of `Fundamental Preferences'{}''
	\textit{Social Choice Welfare} 11: 193--198.
	
\item Kolm, S.-C. (2002) \textit{Justice and Equity} (MIT Press).

\item Koopmans, T.C. (1950) 
	``Utility Analysis of Decisions Affecting Future Well-Being'' (abstract)
	\textit{Econometrica} 18: 174--5.

\item Koopmans, T.C. (1964) ``On Flexibility of Future Preference''
	in M.W. Shelly and G.L. Bryan (eds.) 
	\textit {Human Judgments and Optimality}
	(New York: John Wiley), ch. 13, pp. 243--254.

\item Kreps, D.M. (1990) \textit {Game Theory and Economic Modelling} 
	(Oxford: Clarendon Press).

\item Kreps, D.M. (1992) 
	``Static Choice in the Presence of Unforeseen Contingencies''
	in P.~Dasgupta, D.~Gale, O.~Hart, and E.~Mas\-kin (eds.)
	\textit{Economic Analysis of Markets and Games: 
	Essays in Honor of Frank Hahn\/}
	(Cambridge, Mass.: M.I.T. Press), pp.~258--281.

\item Kreps, D.M., and E.L. Porteus (1978) 
	``Temporal Resolution of Uncertainty and Dynamic Choice Theory''
	\textit{Econometrica} 46 (1): 185--200.

\item Lal, D. (1972) \textit{Wells and Welfare:
	An Exploratory Cost--Benefit Study of Small-Scale Irrigation in Maharashtra}
	(Paris: OECD Development Centre).

\item Little, I.M.D., and J.A. Mirrlees (1969) 
	\textit{Manual of Industrial Project Analysis, vol.\ II} 
	(Paris: OECD Development Centre).

\item Little, I.M.D., and J.A. Mirrlees (1974) 
	\textit{Project Appraisal and Planning} (London: Heinemann).

\item Little, I.M.D., and J.A. Mirrlees (1991) 
	``Project Appraisal and Planning Twenty Years On''
	\textit{Proceedings of the World Bank Annual Conference 
	on Development Economics 1990} pp.\ 351--382.

\item Machina, M. (1989) ``Dynamic Consistency 
	and Non-Expected Utility Models of Choice under Uncertainty''
	\textit{Journal of Economic Literature} 27: 1622--1668.

\item Manzini, P., and M. Mariotti (2007) ``Sequentially Rationalizable Choice''
	\textit{American Economic Review} 97: 1824--1839.

\item Marschak, J. (1950) 
	``Rational Behavior, Uncertain Prospects, and Measurable Utility''
	\textit{Econometrica} 18 (2): 111--141.

\item Metcalfe, S., S. Salles-Filho, L.T. Duarte, A. Bin, A.T. Azevedo, 
	and P.H.A. Feitosa (2021)
	``Shackle's Approach Towards Priority Setting and Decision-Making
	in Science, Technology, and Innovation'' \textit{Futures} 134: 102838.

\item Myerson, R.B. (1983) ``A Dynamic Microeconomic Model 
	with Durable Goods and Adaptive Expectation''
	\textit{Journal of Economic Behaviour and Organization} 4: 309--351.

\item Paul, L.A. (2014) \textit{Transformative Experience} (Oxford U. Press).

\item Paul, L.A. (2015a) ``What You Can't Expect When You're Expecting''
	\textit{Res Philosophica} [online] 92 (2): 1--22.

\item Paul, L.A. (2015b) ``Pr\'ecis of \textit{Transformative Experience}''
	\textit{Philosophy and Phenomenological Research} 91 (3): 760--764.
	
\item Paul, L.A. (2015c) ``Transformative Experience: 
	Replies to Pettigrew, Barnes and Campbell'' 
	\textit{Philosophy and Phenomenological Research} 91 (3): 794--813.
	
\item Pettigrew, R. (2015) ``Transformative Experience and Decision Theory'' 
	\textit{Philosophy and Phenomenological Research} 91 (3): 766--774.

\item Pettigrew, R. (2020) \textit{Choosing for Changing Selves} 
	(Oxford University Press).
	
\item Pivato, M. (2025) ``Universal Recursive Preference Structures'' 	
	available at SSRN: \url{https://ssrn.com/abstract=5410051}.
	
\item Pollak, R.A. (1968) ``Consistent Planning''
	\textit{Review of Economic Studies} 35: 201--208.
	
\item Raiffa, H. (1968) \textit{Decision Analysis: 
	Introductory Lectures on Choices under Uncertainty}
	(Addison-Wesley).
	
\item Randell, P. (2023) ``Familiar Transformative Experiences'' 
	\textit{Synthese} 202: 45.

\item Savage, L.J. (1954, 1972) \textit{Foundations of Statistics}
	(New York: John Wiley; and New York: Dover Publications).
	
\item Schipper, B.C. (2014a) ``Unawareness---A Gentle Introduction 
	to Both the Literature and the Special Issue''
	\textit{Mathematical Social Sciences} 70: 1--9. 	

\item Schipper, B.C. (2014b) ``Preference-Based Unawareness'' 
	\textit{Mathematical Social Sciences} 70: 34--41.
	
\item Schumpeter, J.A. (1911) [2nd edn.\ 1926] 
	\textit{Theorie der wirtschaftlichen Entwicklung; Eine Untersuchung 
	\"uber Unternehmergewinn, Kapital, Kredit, Zins und den Konjunkturzyklus}
	(Leipzig, Duncker \& Humblot).

\item Schumpeter, J.A. (1934) [1961] \textit{The Theory of Economic Development: 
	An Inquiry into Profits, Capital, Credit, Interest, and the Business Cycle}
	translated from Schumpeter (1911) by R. Opie, 
	with a new introduction by J.E. Elliott.

\item Schervish, M.J., T. Seidenfeld, and J.B. Kadane (1990) 
	``State-Dependent Utilities''
	\textit{Journal of the American Statistical Association} 
	85 (411): 840--847.
	
\item Seidenfeld, T., M.J. Schervish, and J.B. Kadane (2010) 
	``Coherent Choice Functions under Uncertainty''	
	\textit{Synthese} 172: 157--176.
	
\item Shackle, G.L.S. (1953) ``The Logic of Surprise''
	\textit{Economica, New Series} 20: 112--117.

\item Shackle, G.L.S. (1969) 
	\textit{Decision, Order, and Time in Human Affairs} 
	(Cambridge: Cambridge University Press). 

\item Silver, D. \textit{et al}.\ (2018) 
	``A General Reinforcement Learning Algorithm 
	that Masters Chess, Shogi, and Go Through Self-Play'' 
	\textit{Science} 362 (issue 6419): 1140--1144.

\item Simon, H.A. (1955) ``A Behavioral Model of Rational Choice''
	\textit{Quarterly Journal of Economics} 69: 99--118.

\item Simon, H.A. (1957) \textit{Models of Man} (New York: John Wiley).

\item Sims, C.A. (2003) ``Implications of Rational Inattention'' 
	\textit{Journal of Monetary Economics} 50 (3): 665--690.
	
\item Sims, C.A. (2011) ``Rational Inattention and Monetary Economics'' 
	\textit{Handbook of Monetary Economics, vol.\ 3} 
	(North-Holland) pp.\ 155--181.

\item Steele, K. (2006) 
	``The Precautionary Principle: A new Approach to Public Decision-making?''
	\textit{Law, Probability and Risk} 5: 19--31.
	
\item Steele, K., and H.O. Stef\'ansson (2021a)
	\textit{Beyond Uncertainty: Reasoning with Unknown Possibilities}
	(Cambridge: Cambridge University Press)

\item Steele, K., and H.O. Stef\'ansson (2021b)	
	``Belief Revision for Growing Awareness''
	\textit{Mind} 130 (520) 1207--1232. 
	
\item Strotz, R.H. (1956) 
	``Myopia and Inconsistency in Dynamic Utility Maximization''
	\textit{Review of Economic Studies} 23: 165--180.

\item Sunstein, C.R. (2003) ``Beyond the Precautionary Principle''
	\textit{University of Pennsylvania Law Review} 151: 1003--1058.

\item Sunstein, C.R. (2005) \textit{Laws of Fear: Beyond the Precautionary Principle}
	(Cambridge University Press: Cambridge).
	
\item Sunstein, C.R., and L.A. Paul (2024) 
	``Freedom, Transformative Experiences, Law, and Testimony'' 
	Harvard Public Law Working Paper No. 24-13. 
	Available at: \url{http://dx.doi.org/10.2139/ssrn.4929159}
	
\item Ullmann-Margalit, E. (2006) 
	``Big Decisions: Opting, Converting, Drifting''
	\textit{Royal Institute of Philosophy Supplement} 58: 157--172.

\item Vier\o, M.-L. (2009) 
	``Exactly What Happens After the Anscombe--Aumann Race? 
	 --- Representing Preferences in Vague Environments''
	 \textit{Economic Theory} 41, 175--212.

\item Vier\o, M.-L. (2021) ``An Intertemporal Model of Growing Awareness''
	\textit{Journal of Economic Theory} 197: 105351.
	
\end{description}
\end{document}